\documentclass[journal,11pt,draftcls,onecolumn]{IEEEtran}

\usepackage{amsmath}
\usepackage{amssymb}
\usepackage{amsthm}
\usepackage{verbatim}
\usepackage{bm}
\usepackage{color,graphicx}
\usepackage{mdwtab}
\usepackage{subfigure}
\usepackage{pstricks}
\usepackage{multirow}
\usepackage{tikz}
\usetikzlibrary{shapes,arrows}

\newtheorem{thm}{Theorem}[]

\newtheorem{ex}{Example}

\newtheorem{prop}[thm]{Proposition}

\def\lightblue{blue!20}
\def\medorange{orange!45}

\def\darkblue{blue!80}
\def\lightorange{orange!20}
\def\darkorange{orange!80}

\def\typical{\mathcal{A}}

\newcommand{\defineqq}{\ensuremath{\stackrel{\text{\tiny def}}{=}}}

\newcommand{\PP}{{\sf Pr}}
\newcommand{\CapR}{\ensuremath{\mathcal C}}

\begin{document}

\bibliographystyle{IEEEtran}

\renewcommand{\textfraction}{0}
\title{Secrecy via Sources and Channels}
\author{\authorblockN{Vinod M. Prabhakaran, Krishnan Eswaran, and Kannan Ramchandran} \\
\authorblockA{
Email: {\tt vinodmp@tifr.res.in}, {\tt krish.eecs@gmail.com}, {\tt kannanr@eecs.berkeley.edu}}
\thanks{V. M. Prabhakaran is with the School of Technology and Computer Science, Tata Institute of Fundamental Research, Mumbai 400005, India. K. Eswaran was with the Department of Electrical Engineering and Computer Sciences, University of California, Berkeley CA 94720, USA; he is now with Google Inc., USA. K. Ramchandran is with the Department of Electrical Engineering and Computer Sciences, University of California, Berkeley CA 94720, USA.}
}
\maketitle

\begin{abstract} Alice and Bob want to share a secret key and to
communicate an independent message, both of which they desire to be kept
secret from an eavesdropper Eve. We study this problem of secret
communication and secret key generation when two resources are available --
correlated sources at Alice, Bob, and Eve, {\em and} a noisy broadcast
channel from Alice to Bob and Eve which is independent of the sources.  We
are interested in characterizing the fundamental trade-off between the
rates of the secret message and secret key.  We present an achievable
solution and prove its optimality for the parallel channels and sources
case when each sub-channel and source component satisfies a degradation
order (either in favor of the legitimate receiver or the eavesdropper).
This includes the case of jointly Gaussian sources and an additive Gaussian
channel, for which the secrecy region is evaluated.
\end{abstract}

\section{Introduction}

Alice has a secret message she wants to send to Bob, but unfortunately, she
must do so in the presence of Eve, an eavesdropper. This paper explores a
new dimension of this familiar problem: how can Alice efficiently utilize
two disparate resources to keep this message secret from Eve? The first
resource is a one-way noisy broadcast channel from Alice to Bob and Eve,
and the second resource is the presence of correlated source observations
at Alice, Bob, and Eve. Specifically, we are interested in understanding
how to design strategies that ``fuse'' these resources optimally in order
to support secure communication between Alice and Bob.
 
There already exists a body of literature for cases in which only one of
these resources is available. Wyner's seminal work, ``The Wire-tap
Channel''~\cite{Wyner75wtc} considered secure communication over degraded
broadcast channels~\cite{CoverBC} and was later generalized by Csisz\'{a}r
and K\"{o}rner~\cite{Csiszar78bcc} to cover all broadcast channels.

Analogously, Ahlswede and Csisz\'{a}r~\cite{AhlswedeCSecret93} and
Maurer~\cite{Maurer93} recognized that dependent source observations
available at the terminals can be used as a resource for generating a
secret-key -- a uniform random variable shared by Alice and Bob which Eve is
oblivious of -- if the terminals can communicate over a noiseless public
channel (which delivers all its input faithfully to all the terminals
including the eavesdropper). In~\cite{AhlswedeCSecret93}, the secret-key
capacity of dependent sources was characterized if a one-way noiseless
public channel from Alice to Bob and Eve of unconstrained capacity is
available. The characterization for the case when there is a constraint on
the capacity of the public channel was later found by Csisz\'{a}r and
Narayan~\cite{CsiszarNarayan2000} as a special case of their results on a
class of common randomness generation problems using a helper.  As in the
channel setting, one can also exploit distributed sources for sending a
secret message.

The present investigation is motivated by wireless sensor networks, in
which sensors have access to both a wireless channel and their correlated
sensor readings. Note that in such situations, fading can cause the channel
characteristics to be more or less favorable to secrecy at different points
in time. Thus, when the channel characteristics are favorable, it can be
advantageous for Alice and Bob, instead of (or in addition to) sending a
specific secret message, to simply agree on a sequence of private common
random bits (a secret key) to be used later when the characteristics are
unfavorable. See Khalil et al. \cite{khalil-2009} for an example of how
this can enable a form of secure communication with delay constraints under
fading channels. Not surprisingly, it turns out that in some settings, one
can achieve higher rates for the secret key than the more restrictive
secret message. The general problem we consider abstracts this issue into
considering a tradeoff between transmitting a uniform source privately (a
secret message) and generating private common randomness (a secret key).  A
related model for the secret message case was studied by Chen and Vinck
\cite{ChenVinck}, who consider a channel with non-causal channel state
information at Alice, and the channel is degraded in favor of Bob, as in
Wyner's wiretap channel. A more recent work by Khisti, Diggavi and
Wornell~\cite{KhistiDiggaviWornell11} also examines secret key agreement
when non-causal channel state information is available at Alice. The
achievability result there coincides with the results in this paper when it
is specialized to the one there. In independent and concurrent work as
ours, Khisti, Diggavi and Wornell~\cite{KhistiDiggaviWornell12} also
investigate secrecy in a similar setting, but their focus is solely on the
question of secret key generation and limited to the case where only Alice
and Bob have source observations.  The achievability results and an
optimality result in their work coincide with the results in this paper
under the specialized setting above. Additionally, a general upper bound
on the secret key capacity is provided in~\cite{KhistiDiggaviWornell12}.

In contradistinction to these other works, our main contributions are (i) an
achievable trade-off between secret-key and secret-message rates when both
dependent sources and a one-way broadcast channel are available, (ii) a proof
of optimality of this trade-off for parallel channels and sources when each
sub-channel and source component satisfies a degradation order either in favor
of Bob or in favor or of Eve, and (iii) evaluation of this optimal trade-off in
the Gaussian case.

Section \ref{sect:secrecy_setup} gives a formal description of the problem
setup, and Section \ref{sect:secrecy_results} describes the main results
presented in this work.  Section~\ref{sect:secrecy_examples} gives an
interpretation to the achievability part of the coding theorem The paper
concludes with a discussion and directions for future work in Section
\ref{sect:secrecy_discussion}.

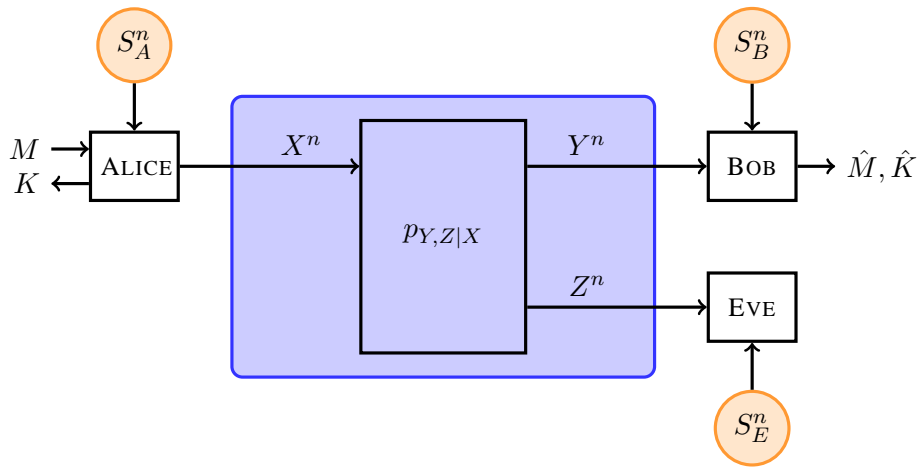
\begin{figure}[ht]
	\centering
        \pgfdeclarelayer{background}
        \pgfdeclarelayer{foreground}
        \pgfsetlayers{background,main,foreground}

        \tikzstyle{channelwrap} = [text width=6em, fill=blue!20, 
            minimum height=12em, rounded corners]
        \tikzstyle{channel}=[draw, text width=5em, 
            text centered, minimum height=8em]
        \tikzstyle{party}=[draw, text width=2.35em, 
            text centered, minimum height=2.35em]
        \tikzstyle{source}=[circle, draw=orange!80, fill=orange!20,
            text centered, minimum size=.6em]

        \def\blockdist{2.3}
        \def\edgedist{.5}
        \begin{tikzpicture}

        \begin{scope}[very thick] 
            \node (channel) [channel] {$p_{Y,Z|X}$};
            \path (channel.40)+(1.3*\blockdist,0) node (bob) [party] {\small \sc{Bob}};
            \path (channel.-40)+(1.3*\blockdist,0) node (eve) [party] {\small \sc{Eve}};
            \path (channel.140)+(-1.3*\blockdist,0) node (alice) [party] {\small \sc{Alice}};
            \path[draw, ->] (channel.east |- bob) -- node (yn) [above left] {$Y^n$} (bob);
            \path[draw, ->] (channel.east |- eve) -- node (zn) [above left] {$Z^n$} (eve);
            \path[draw, ->] (alice) -- node (xn) [above right] {$X^n$} (alice -| channel.west);
            \draw[->] (bob.east) --  node {} + (\edgedist,0) node[right] {$\hat{M}, \hat{K}$};
            \draw[->] (alice.west)+(0,-0.1*\blockdist) --  node {} + (-\edgedist,-0.1*\blockdist) node[left] {${K}$};
            \draw[<-] (alice.west)+(0,0.1*\blockdist) --  node {} + (-\edgedist,0.1*\blockdist) node[left] {${M}$};

            \path (alice)+(0,.7*\blockdist) node (asource) [source] {$S_A^n$};
            \path (bob)+(0,.7*\blockdist) node (bsource) [source] {$S_B^n$};
            \path (eve)+(0,-.7*\blockdist) node (esource) [source] {$S_E^n$};
            \path[draw, ->] (asource.south) -- (alice.north);
            \path[draw, ->] (bsource.south) -- (bob.north);
            \path[draw, ->] (esource.north) -- (eve.south);

            \begin{pgfonlayer}{background}
               \path (xn.west |- channel.north)+(-.5,.3) node (a) {};
               \path (channel.south -| zn.east)+(.5,-.3) node (b) {};
               \path[fill=blue!20,rounded corners, draw=blue!80, very thick]
                    (a) rectangle (b);
            \end{pgfonlayer}
               
        \end{scope}

        \end{tikzpicture}
	\caption{Problem setup: Alice and Bob want to share a key $K$ and
independent message $M$, both of which they want to be kept secret from Eve.
Alice has a memoryless broadcast channel to Bob and Eve. Additionally, Alice, Bob, and Eve have make correlated memoryless source observations.}
\end{figure}

\section{Problem Setup \label{sect:secrecy_setup}}

\noindent{\em Notation:} We denote random variables by upper-case letters
({\em e.g.}, $X$), their realizations by lower-case letters ({\em e.g.},
$x$), and the alphabets over which they take values by calligraphic
letters ({\em e.g.}, ${\mathcal X}$). A vector $(X_k, X_{k+1},\ldots,X_n)$
will be denoted by $X_k^n$. When $k=1$, the subscript will be dropped  as
in $X^n=(X_1,X_2,\ldots,X_n)$.

We consider the following model. Alice, Bob and Eve observe, respectively,
the dependent memoryless processes (sources) $S_{A,k}, S_{B,k}, S_{E,k}$,
where $k=1,2,\ldots$ is the time index. They have a joint distribution
$p_{S_A,S_B,S_E}$ over the alphabet ${\mathcal S}_A\times{\mathcal
S}_B\times{\mathcal S}_E$. Independent of these sources, there is a
memoryless broadcast channel from Alice to Bob and Eve given by
$p_{Y,Z|X}$, where $X_k$ is the input to the channel, $Y_k$ is Bob's
output, and $Z_k$ Eve's. We will also allow Alice to have access to a
private random variable $\Phi_A$ which is not available to Bob and Eve and
which is independent of all other random variables. Alice may use this
private random variable for purposes of randomization.

For $\epsilon>0$, a random variable $U$ is defined to be
{\em$\epsilon$-recoverable} from another random variable $V$ if there is a
function $f$ such that $\text{Pr}(U\neq f(V))\leq\epsilon$.  Suppose the
parties make $n$ observations of their sources, and Alice sends an
$n$-length input $X^n$ to the channel. The input is a function of the
observation $S_A^n$, the secret message $M$ which is uniformly distributed
over its alphabet ${\mathcal M}$ and independent of the sources and
channel, and the private random variable $\Phi_A$ available only to Alice.
We say that $K=g(S_A^n,\Phi_A)$, for some $g$, is an
{\em$\epsilon$-secret-key} if (i) it is $\epsilon$-recoverable from
$S_B^n,Y^n$, (ii) satisfies the secrecy condition\footnote{A stronger form
of secrecy can be achieved by directly invoking the ideas in~\cite{MaurerWolf00} as
will be briefly discussed in Section~\ref{sect:secrecy_discussion}.}
\begin{align}
\frac{1}{n}I(M,K;Z^n,S_E^n) \leq \epsilon, \label{eq:secrecy_cond}
\end{align} and (iii) satisfies the
uniformity condition \[\frac{1}{n}H(K) \geq \frac{1}{n}\log |{\mathcal K}|
- \epsilon,\] where ${\mathcal K}$ is the alphabet over which $K$ takes its
values.  We define $(R_{\sf SK,\epsilon},R_{\sf SM,\epsilon})$ to be an
{\em $\epsilon$-achievable rate pair} if there is an $\epsilon$-secret-key
$K^n$ such that $\frac{1}{n}H(K^{(n)})=R_{\sf SK,\epsilon}$, the secret
message $M$ is $\epsilon$-recoverable from $(Y^n,S_B^n)$, and $\frac{1}{n}
\log |{\mathcal M}|=R_{\sf SM,\epsilon}$. A rate pair $(R_{\sf SK},R_{\sf
SM})$ is said to be {\em achievable} if there is a sequence of $\epsilon_n$
such that $(R_{\sf SK,\epsilon_n},R_{\sf SM,\epsilon_n})$ are
$\epsilon_n$-achievable rate pairs, and as $n\rightarrow\infty$, \[
\epsilon_n\rightarrow 0, \qquad R_{\sf SK,\epsilon_n}\rightarrow R_{\sf
SK}, \qquad\text{ and }\qquad R_{\sf SM,\epsilon_n}\rightarrow R_{\sf
SM}.\] We define the rate region $\CapR$ to be the set of all
achievable rate pairs.

\section{Results \label{sect:secrecy_results}}
In order to state our main results we will consider a more general setup than what we described above.
Consider a memoryless broadcast channel $p_{\mathbf{Y},\mathbf{Z}|X,S}$ with
non-causal state information $S^n$ available at the encoder Alice whose input to the channel is $X^n$; Bob and Eve receive, respectively, $\mathbf{Y}^n$ and $\mathbf{Z}^n$. The state
sequence $S^n$ is independent and identically distributed with a
probability mass function $p_S$.
Note that the setting in Section~\ref{sect:secrecy_setup} is a special case
with $S_k=S_{A,k}$, $\mathbf{Y}_k = (Y_k,S_{B,k})$, and 
$\mathbf{Z}_k=(Z_k,S_{E,k})$. 

Let $\mathcal{P}_{\sf joint}$ be the set of all joint distributions
$p$ of random variables $\mathbf{V},\mathbf{U},X,S,\mathbf{Y}, \mathbf{Z}$
such that (i) the following Markov chain holds:
		\begin{align*}
			\mathbf{V} - \mathbf{U} - (X,S) - (\mathbf{Y}, \mathbf{Z})~,
		\end{align*}
(ii) $\mathbf{V}$ is independent of $S$, and (iii) the joint conditional
distribution of $(\mathbf{Y}, \mathbf{Z})$ given $(X,S)$ as well as the
marginal distribution of $S$ are consistent with the given source and
channel respectively. Let $\CapR_{\sf joint}$ denote the set of all achievable rate pairs 
for this channel.
	
For $p \in \mathcal{P}_{\sf joint}$, let ${\mathcal R}_{\sf joint}(p)$ be the set of all non-negative pairs $(R_{\sf SK}, R_{\sf SM})$ which satisfy the following two inequalities:
		\begin{align}
			R_{\sf SM} &\leq I(\mathbf{U}; \mathbf{Y}) - I(\mathbf{U};S) \\
			R_{\sf SK}+R_{\sf SM} &\leq I(\mathbf{U}; \mathbf{Y} | \mathbf{V}) - I(\mathbf{U};\mathbf{Z} | \mathbf{V})~.
		\end{align}
We prove the following theorem in Appendix~\ref{app:achievabilityproof}.	
\begin{thm} \label{thm:joint_encoding}
		\begin{align}
			\CapR_{\sf joint} \supseteq \bigcup_{p \in \mathcal{P}_{\sf joint}} {\mathcal R}_{\sf joint}(p)~.
		\end{align} 
\end{thm}

We obtain our main achievability result as a corollary of the above theorem.

Let ${\mathcal P}$ be the set of all joint distributions $p$ of random
variables $U_1,V_1,V_2,X,Y,Z,S_A,S_B,S_E$ such that (i) $(U_1,S_A,S_B,S_E)$ and $(V_1,V_2,X,Y,Z)$
are independent, (ii) the following two Markov chains hold:
\begin{align*}
U_1 - S_A &- (S_B,S_E),\\
V_2 - V_1 -& X - (Y,Z),
\end{align*}
(iii) the joint distribution of $(S_A,S_B,S_E)$ and the joint conditional
distribution of $(Y,Z)$ given $X$ are consistent with the given source and
channel respectively, and (iv) the following inequality holds:
\begin{align}
 I(V_1;Y) \geq I(U_1;S_A) - I(U_1;S_B). \label{eq:pcondition}
\end{align}
For $p\in{\mathcal P}$, let ${\mathcal R}(p)$ be the set of all 
non-negative pairs $(R_{\sf SK},R_{\sf SM})$ which satisfy
the following two inequalities 
\begin{align}
R_{\sf SM} &\leq I(V_1;Y) - ( I(U_1 ; S_A) -  I(U_1;S_B) ),\label{eq:inner1}\\
R_{\sf SK}  + R_{\sf SM} &\leq [I(V_1;Y|V_2) - I(V_1;Z|V_2)]_+ 
         + [I(U_1;S_B) - I(U_1;S_E)]_+,\label{eq:inner2}
\end{align}
where $[x]_+\defineqq\max(0,x)$. The next theorem states that all pairs of
rates belonging to ${\mathcal R}(p)$ are achievable. An interpretation of 
the result is presented in Section~\ref{sect:secrecy_examples}.
\begin{thm}\label{thm:achievability}
\[\CapR \supseteq \bigcup_{p\in{\mathcal P}} {\mathcal R}(p).\]
\end{thm}
\noindent{\em Remark:} It can be shown that in taking the union above, it
suffices to consider auxiliary random variables with a sufficiently large, but
finite cardinality. In particular, we may restrict the sizes of the alphabets
${\mathcal U}_1, {\mathcal V}_1, {\mathcal V}_2$ of the auxiliary random
variables $U_1,V_1,V_2$, respectively, to $|{\mathcal U}_1|=|{\mathcal
S}_A|+2$, $|{\mathcal V}_1| = (|{\mathcal X}|+3)(|{\mathcal X}|+1)$ and
$|{\mathcal V}_2|=|{\mathcal X}|+3$.  This can be shown using a strengthened
form of Fenchel-Eggleston-Carath\'{e}odery's theorem~\cite[pg.
310]{CsiszarKornerbook1ed} (see, for example,~\cite{Csiszar78bcc} for a similar
calculation).

\noindent\textit{Proof of Theorem \ref{thm:achievability}}.
\noindent Set $\mathbf{V} = V_2$ and $\mathbf{U} = (U_1, V_1)$ in Theorem~\ref{thm:joint_encoding}.
Then we have the following:
	\begin{align*}
		I(\mathbf{U}; \mathbf{Y}) - I(\mathbf{U};S) &= I(V_1 ; Y) + I(U_1 ; S_B) - I(U_1 ;S_A) \\
		I(\mathbf{U}; \mathbf{Y} | \mathbf{V}) - I(\mathbf{U};\mathbf{Z} | \mathbf{V}) &= I(V_1 ; Y | V_2) - I(V_1; Z | V_2) + I(U_1 ; S_B) - I(U_1 ; S_E)
	\end{align*}
Note that if $I(U_1 ; S_B) - I(U_1 ; S_E) \leq 0$, we can increase the achievable region by making $U_1$ independent of $S_A$. Likewise, if $I(V_1 ; Y | V_2) - I(V_1; Z | V_2) < 0$, we can increase the region 
by making $V_1 = V_2$. Thus, we have established the rate region in Theorem \ref{thm:achievability} 
as a special case.
\qed

The next theorem states that the above inner bound is tight for the case of
parallel channels and sources where each sub-channel and source component
satisfies a degradation order (either in favor of the legitimate receiver
or in favor of the eavesdropper).

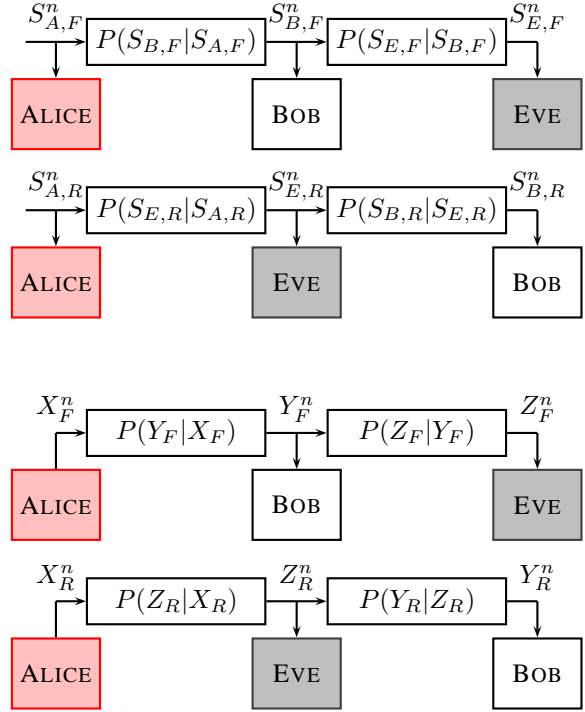
\begin{figure}[ht]
	\centering
	\vspace{1cm}
        \begin{pspicture}(-0.5,-1)(5,7)
        \psset{unit=1.6cm}
        \rput(0,0){\rput(.75,.2){\rput(0,0){\small{$X_R^n$}}
                \rput(0,-0.2){\psline(0,0)(0,-0.3)}
                \rput(-0.5,-0.8){\psframe[linecolor=red, fillstyle=solid, fillcolor=pink](0.125,-0.3)(.875,0.3)} %
                \rput(0,-0.8){\small{\sc{Alice}}}%
                }
                \rput(.75,0){\psline{->}(0,0)(.25,0)}

                \rput(1,0){\psframe(0,-0.2)(1.5,0.2)} %
                \rput(1.75,0){\small{$P(Z_R|X_R)$}}%

                \rput(2.75,0.2){\rput(0,0){\small{$Z_R^n$}}
                \rput(0,-0.2){\psline{->}(0,0)(0,-0.3)}
                \rput(-0.5,-0.8){\psframe[linecolor=gray!50!black, fillstyle=solid, fillcolor=gray!50!white](0.125,-0.3)(.875,0.3)} %
                \rput(0,-0.8){\small{\sc{Eve}}}%
                }
                \rput(2.5,0){\psline{->}(0,0)(0.5,0)}

                \rput(3,0){\psframe(0,-0.2)(1.5,0.2)} %
                \rput(3.75,0){\small{$P(Y_R|Z_R)$}}%

                \rput(4.75,0.2){\rput(0,0){\small{$Y_R^n$}}
                \rput(0,-0.2){\psline{->}(0,0)(0,-0.3)}
                \rput(-0.5,-0.8){\psframe[linecolor=black, fillstyle=solid, fillcolor=white](0.125,-0.3)(.875,0.3)} %
                \rput(0,-0.8){\small{\sc{Bob}}}%
                }
                \rput(4.5,0){\psline(0,0)(0.25,0)}
        }

        \rput(0,1.4){\rput(.75,.2){\rput(0,0){\small{$X_F^n$}}
                \rput(0,-0.2){\psline(0,0)(0,-0.3)}
                \rput(-0.5,-0.8){\psframe[linecolor=red, fillstyle=solid, fillcolor=pink](0.125,-0.3)(.875,0.3)} %
                \rput(0,-0.8){\small{\sc{Alice}}}%
                }
                \rput(.75,0){\psline{->}(0,0)(.25,0)}

                \rput(1,0){\psframe(0,-0.2)(1.5,0.2)} %
                \rput(1.75,0){\small{$P(Y_F|X_F)$}}%

                \rput(2.75,0.2){\rput(0,0){\small{$Y_F^n$}}
                \rput(0,-0.2){\psline{->}(0,0)(0,-0.3)}
                \rput(-0.5,-0.8){\psframe[linecolor=black, fillstyle=solid, fillcolor=white](0.125,-0.3)(.875,0.3)} %
                \rput(0,-0.8){\small{\sc{Bob}}}%
                }
                \rput(2.5,0){\psline{->}(0,0)(0.5,0)}

                \rput(3,0){\psframe(0,-0.2)(1.5,0.2)} %
                \rput(3.75,0){\small{$P(Z_F|Y_F)$}}%

                \rput(4.75,0.2){\rput(0,0){\small{$Z_F^n$}}
                \rput(0,-0.2){\psline{->}(0,0)(0,-0.3)}
                \rput(-0.5,-0.8){\psframe[linecolor=gray!50!black, fillstyle=solid, fillcolor=gray!50!white](0.125,-0.3)(.875,0.3)} %
                \rput(0,-0.8){\small{\sc{Eve}}}%
                }
                \rput(4.5,0){\psline(0,0)(0.25,0)}

        }

        \rput(0,4.65){\rput(.75,.2){\rput(0,0){\small{$S_{A,F}^n$}}
                \rput(0,-0.2){\psline{->}(0,0)(0,-0.3)}
                \rput(-0.5,-0.8){\psframe[linecolor=red, fillstyle=solid, fillcolor=pink](0.125,-0.3)(.875,0.3)} %
                \rput(0,-0.8){\small{\sc{Alice}}}%
                }
                \rput(.5,0){\psline{->}(0,0)(.5,0)}

                \rput(1,0){\psframe(0,-0.2)(1.5,0.2)} %
                \rput(1.75,0){\small{$P(S_{B,F}|S_{A,F})$}}%

                \rput(2.75,0.2){\rput(0,0){\small{$S_{B,F}^n$}}
                \rput(0,-0.2){\psline{->}(0,0)(0,-0.3)}
                \rput(-0.5,-0.8){\psframe[linecolor=black, fillstyle=solid, fillcolor=white](0.125,-0.3)(.875,0.3)} %
                \rput(0,-0.8){\small{\sc{Bob}}}%
                }
                \rput(2.5,0){\psline{->}(0,0)(0.5,0)}

                \rput(3,0){\psframe(0,-0.2)(1.5,0.2)} %
                \rput(3.75,0){\small{$P(S_{E,F}|S_{B,F})$}}%

                \rput(4.75,0.2){\rput(0,0){\small{$S_{E,F}^n$}}
                \rput(0,-0.2){\psline{->}(0,0)(0,-0.3)}
                \rput(-0.5,-0.8){\psframe[linecolor=gray!50!black, fillstyle=solid, fillcolor=gray!50!white](0.125,-0.3)(.875,0.3)} %
                \rput(0,-0.8){\small{\sc{Eve}}}%
                }
                \rput(4.5,0){\psline(0,0)(0.25,0)}

        }

        \rput(0,3.25){\rput(.75,.2){\rput(0,0){\small{$S_{A,R}^n$}}
                \rput(0,-0.2){\psline{->}(0,0)(0,-0.3)}
                \rput(-0.5,-0.8){\psframe[linecolor=red, fillstyle=solid, fillcolor=pink](0.125,-0.3)(.875,0.3)} %
                \rput(0,-0.8){\small{\sc{Alice}}}%
                }
                \rput(.5,0){\psline{->}(0,0)(.5,0)}

                \rput(1,0){\psframe(0,-0.2)(1.5,0.2)} %
                \rput(1.75,0){\small{$P(S_{E,R}|S_{A,R})$}}%

                \rput(2.75,0.2){\rput(0,0){\small{$S_{E,R}^n$}}
                \rput(0,-0.2){\psline{->}(0,0)(0,-0.3)}
                \rput(-0.5,-0.8){\psframe[linecolor=gray!50!black, fillstyle=solid, fillcolor=gray!50!white](0.125,-0.3)(.875,0.3)} %
                \rput(0,-0.8){\small{\sc{Eve}}}%
                }
                \rput(2.5,0){\psline{->}(0,0)(0.5,0)}

                \rput(3,0){\psframe(0,-0.2)(1.5,0.2)} %
                \rput(3.75,0){\small{$P(S_{B,R}|S_{E,R})$}}%

                \rput(4.75,0.2){\rput(0,0){\small{$S_{B,R}^n$}}
                \rput(0,-0.2){\psline{->}(0,0)(0,-0.3)}
                \rput(-0.5,-0.8){\psframe[linecolor=black, fillstyle=solid, fillcolor=white](0.125,-0.3)(.875,0.3)} %
                \rput(0,-0.8){\small{\sc{Bob}}}%
                }
                \rput(4.5,0){\psline(0,0)(0.25,0)}

        }
        \end{pspicture}
	\vspace{.2cm}
	\caption{Theorem \ref{thm:outerbounds} states that the inner bound to the rate region $\CapR$ established in Theorem \ref{thm:achievability} is tight if the sources and channels can be decomposed to satisfy a degradation order, either in favor of Bob or Eve.}
\end{figure}

\begin{thm}\label{thm:outerbounds}
Consider the following:
\begin{itemize}
\item[(i)] The channel has two independent components\footnote{We denote the
channel input, outputs, and the sources using bold letters to make this
explicit.} denoted by $F$ and $R$: ${\bf X}=(X_F,X_{R})$, ${\bf Y}=(Y_F,Y_{R})$, and ${\bf Z}=(Z_F,Z_{R})$
such that
$p_{Y_F,Y_{R},Z_F,Z_{R}|X_F,X_{R}}=p_{Y_F,Z_F|X_F}p_{Y_{R},Z_{R}|X_{R}}$.
Moreover, the first sub-channel $F$ is degraded in favor of Bob, which we call 
forwardly degraded, and the
second sub-channel $R$ is degraded in favor of Eve, which we 
call reversely degraded; {\em i.e.}, $X_F-Y_F-Z_F$
and $X_{R}-Z_{R}-Y_{R}$ are Markov chains.
\item[(ii)] The sources also have two independent components, again denoted
by $F$ and $R$: ${\bf S}_A=(S_{A,F},S_{A,R})$, ${\bf S}_B=(S_{B,F},S_{B,R})$,
and ${\bf S}_E=(S_{E,F},S_{E,R})$ with
$p_{{\bf S}_A,{\bf S}_B,{\bf S}_E}=p_{S_{A,F},S_{B,F},S_{E,F}}p_{S_{A,R},S_{B,R}S_{E,R}}$.
The first component is degraded in favor of Bob and the second in favor of
Eve; {\em i.e.}, $S_{A,F}-S_{B,F}-S_{E,F}$ and $S_{A,R}-S_{E,R}-S_{B,R}$
are Markov chains.
\end{itemize}

In this case,
\[ \CapR = \bigcup_{p\in\tilde{\mathcal P}}\tilde{\mathcal R}(p),\]
where $\tilde{\mathcal P}$ is the set of joint distributions of the form
$p_{V_{2,F},X_F}p_{Y_F,Z_F|X_F}$$p_{X_R}p_{Y_R,Z_R|X_R}$$p_{U_{1,F}|S_{A,F}}
p_{S_{A,F},S_{B,R},S_{E,R}}$ $p_{S_{A,R},S_{B,R},S_{E,R}}$
and $\tilde{\mathcal R}(p)$ is the set of non-negative pairs of $(R_{\sf
SK},R_{\sf SM})$ satisfying 
\begin{align}
R_{\sf SM} &\leq I(X_F;Y_F ) + I(X_{R};Y_{R} ) - (I(U_{1,F};S_{A,F}) -
I(U_{1,F};S_{B,F})),
\text{ and} \label{eq:casea1}\\
R_{\sf SK}  + R_{\sf SM} &\leq I(X_F;Y_F|V_{2,F}) - I(X_F;Z_F|V_{2,F}) +
                               I(U_{1,F};S_{B,F}) - I(U_{1,F};S_{E,F}).\label{eq:casea2}
\end{align}
\end{thm}

We prove this theorem in Appendix~\ref{app:outerbounds} where we also show
that, as one would expect, the result holds even if we only have stochastic
degradation instead of physical degradation. It turns out the result is more
general than the form presented above, but these extensions are omitted to
be able to state the result cleanly. These extensions are discussed in
greater detail in Section \ref{sect:secrecy_discussion}.

\begin{figure}[htb]
\begin{center}
\pgfdeclarelayer{background}
\pgfdeclarelayer{foreground}
\pgfsetlayers{background,main,foreground}

\tikzstyle{channelwrap} = [text width=6em, fill=blue!20, 
    minimum height=12em, rounded corners]
\tikzstyle{channel}=[draw, text width=5em, 
    text centered, minimum height=8em]
\tikzstyle{party}=[draw, text width=2.35em, 
    text centered, minimum height=2.35em]
\tikzstyle{source}=[circle, draw=orange!80, fill=orange!20,
    text centered, minimum size=.6em]
\tikzstyle{adder}=[circle, text centered, draw=black, minimum size=.2em, very thick]

\def\blockdist{2.3}
\def\edgedist{.5}
\begin{tikzpicture}

\begin{scope}[very thick] 
    \node (bobchadder) [adder] {\bf +};
    \path (bobchadder.south)+(0,-1.5*\blockdist) node (evechadder) [adder] {\bf +};

    \path (bobchadder)+(1.3*\blockdist,0) node (bob) [party] {\small \sc{Bob}};
    \path (evechadder)+(1.3*\blockdist,0) node (eve) [party] {\small \sc{Eve}};
    \path (bobchadder)+(-1.3*\blockdist,0) node (alice) [party] {\small \sc{Alice}};
    \path[draw, ->] (bobchadder.east) -- node (yn) [above left] {$Y^n$} (bob);
    \path[draw, ->] (evechadder.east |- eve) -- node (zn) [above left] {$Z^n$} (eve);
    \path[draw, ->] (alice) -- node (xn) [above right] {$X^n$} (alice -| bobchadder.west);
    \draw[->] (bob.east) --  node {} + (\edgedist,0) node[right] {$\hat{M}, \hat{K}$};
    \draw[->] (alice.west)+(0,-0.1*\blockdist) --  node {} + (-\edgedist,-0.1*\blockdist) node[left] {${K}$};
    \draw[<-] (alice.west)+(0,0.1*\blockdist) --  node {} + (-\edgedist,0.1*\blockdist) node[left] {${M}$};

    \path (alice)+(0,.7*\blockdist) node (asource) [source] {$S_A^n$};
    \path (bob)+(0,.7*\blockdist) node (bsource) [source] {$S_B^n$};
    \path[draw, ->] (asource.south) -- (alice.north);
    \path[draw, ->] (bsource.south) -- (bob.north);

    \path (xn|-bobchadder) node (taptop) [above] {};
    \path (taptop |- evechadder) node (tapbottom) [below] {};
    \path[draw,-] (taptop) -- (tapbottom);
    \path[draw,->] (tapbottom.north) -- (evechadder.west);

    \path (bobchadder|-asource) node (bobsrcadder) [adder] {\bf +};
    \path[draw, ->] (asource.east) -- (bobsrcadder.west);
    \path[draw, ->] (bobsrcadder.east) -- (bsource.west);
    \draw[<-] (bobsrcadder.north) -- node {} + (0,\edgedist) node[above] {$N_\text{source}$};    
    \draw[<-] (evechadder.north) -- node {} + (0,\edgedist) node[above] {$N_\text{Eve}$};    
    \draw[<-] (bobchadder.south) -- node {} + (0,-\edgedist) node[below] {$N_\text{Bob}$};    
 
    \begin{pgfonlayer}{background}
       \path (xn.west |- bobchadder.north)+(-.5,.3) node (a) {};
       \path (evechadder.south -| zn.east)+(.5,-.3) node (b) {};
       \path[fill=blue!20,rounded corners, draw=blue!80, very thick]
            (a) rectangle (b);
    \end{pgfonlayer}
 
\end{scope}

\end{tikzpicture}
\end{center}
\caption{
The scalar Gaussian case with no source observations at Eve.
}\label{fig:Gaussian_example}
\end{figure}
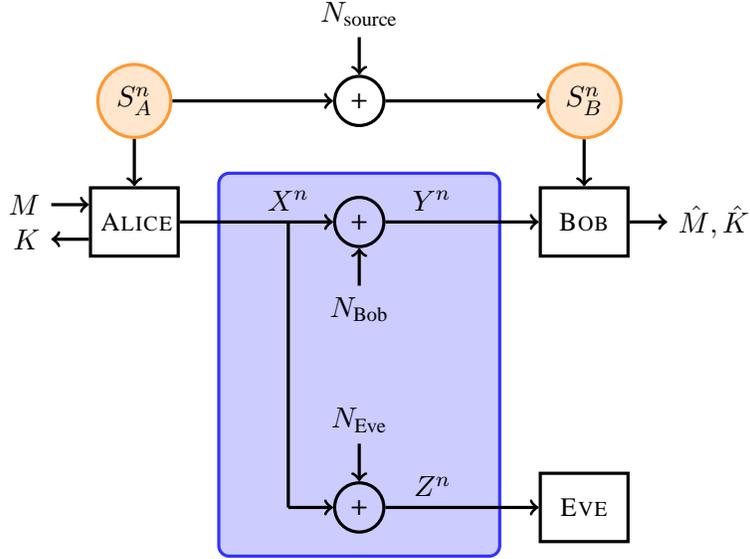

\begin{figure}[ht]
\begin{center}
	\includegraphics[width=15cm]{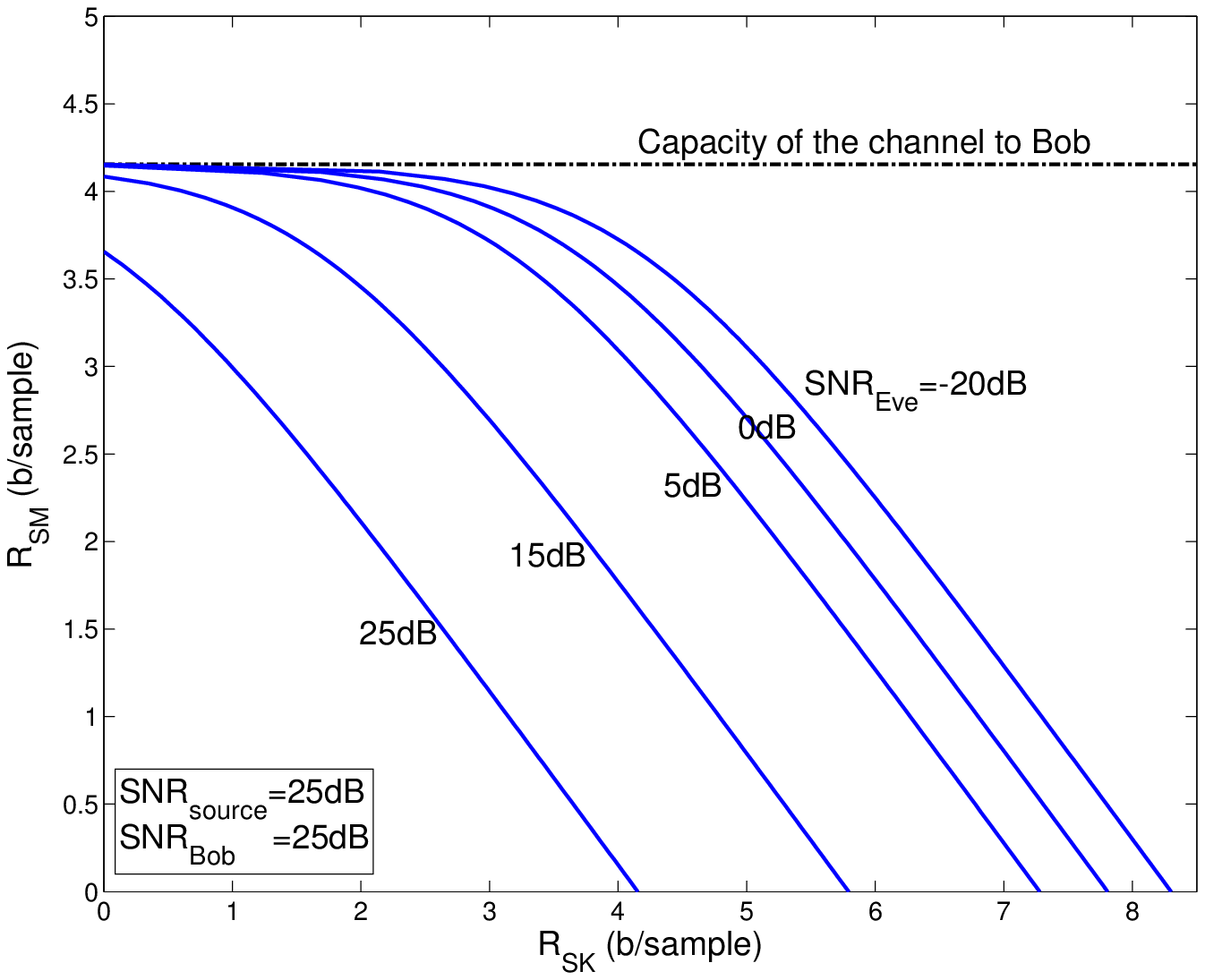}
\end{center}
\caption{The figure plots the optimal tradeoff between secret key and secret message from Proposition 
\ref{prop:Gaussianexample}, which is the special case in which there is no source at Eve. The tradeoff 
curves above reveal distinguishing features between the secret key and secret message rates. For instance, the largest possible secret key rate is greater than that of the largest possible secret message rate, and the tradeoff between key and message is governed by a curve that is not simply linear. \label{fig:Gaussian_tradeoff_curve}}
\end{figure}
\subsection{The scalar Gaussian case}\label{sec:gaussian_example}
Let us consider a scalar Gaussian example (Figure~\ref{fig:Gaussian_example}). Suppose the observations of
Alice and Bob are jointly Gaussian. Then, without loss of generality, we
can model them as
\[ S_B = S_A + N_\text{source},\]
where $S_A$ and $N_\text{source}$ are independent zero mean Gaussian. Let
$N_\text{source}$ be unit variance, and let the variance of $S_A$ be
$\text{SNR}_\text{src}$. Let Eve have no source observation. Suppose
that the broadcast channel has additive Gaussian noise with a 
power constraint on $X$ of $\text{SNR}_\text{Bob}$. Let
\begin{align*}
Y&=X+N_\text{Bob},\text{ and}\\
Z&=X+N_\text{Eve},
\end{align*}
where $N_\text{Bob}$ and $N_\text{Eve}$ are
Gaussians independent of $X$, and such that $N_\text{Bob}$ has unit
variance and $N_\text{Eve}$ has a variance
$\text{SNR}_\text{Bob}/\text{SNR}_\text{Eve}$.
We have the following proposition, which is plotted in Figure \ref{fig:Gaussian_tradeoff_curve} 
and proved in
Appendix~\ref{app:Gaussianexample}.
\begin{prop} \label{prop:Gaussianexample}
The rate region $\CapR$ for this problem is set of all
non-negative $(R_{\sf SK},R_{\sf SM})$ pairs satisfying
\begin{align*}
R_{\sf SM} &\leq \frac{1}{2}\log
\frac{(1+\text{SNR}_\text{src})(1+\text{SNR}_\text{Bob})}
{1+\text{SNR}_\text{src}+\min(\text{SNR}_\text{Bob},\text{SNR}_\text{Eve})},
\\
R_{\sf SK}&\leq \frac{1}{2}\log
\frac{(1+\text{SNR}_\text{src})(1+\text{SNR}_\text{Bob})
\exp(-2R_{\sf SM}) - \text{SNR}_\text{src}}
{1+\min(\text{SNR}_\text{Bob},\text{SNR}_\text{Eve})}
\end{align*}
\end{prop}

\noindent{\em Remark:} When Eve also has a source observation jointly
Gaussian with the observations of Alice and Bob, the problem is covered by
the cases in Theorem~\ref{thm:outerbounds}. However, unlike in the
proposition above, we were unable to show that a Gaussian choice of the
auxiliary random variables is optimal. Indeed, even for the secret key
problem under jointly Gaussian sources and only a public bit-pipe channel
from Alice to Bob and Eve, the optimality of Gaussian auxiliary random
variables remains open to the best of our knowledge.

\section{Intuition behind Theorem~\ref{thm:achievability}: A separation strategy \label{sect:secrecy_examples}}

In this section we will sketch informally the intuition behind the
achievable scheme of Theorem~\ref{thm:achievability}. We will briefly
describe three examples before proceeding. Examples \ref{ex:intro_ch} and
\ref{ex:intro_src} highlight well known achievable strategies in the
secrecy literature. The key idea is shown in Example \ref{ex:intro};
namely, that the strategies in Examples \ref{ex:intro_ch} and
\ref{ex:intro_src} can be used as building blocks to construct a strategy
for Example \ref{ex:intro}, much in the same way a source and a channel
code can be used as building blocks to construct an achievable strategy in
a joint source-channel context. This is what we mean by a separation
strategy, which establishes the basic intuition for
Theorem~\ref{thm:achievability}. The remainder of the section extends this
to the more general problem setup of the paper.

\begin{ex} \label{ex:intro_ch}
Suppose Alice has a three-bit noiseless channel $(x_1, x_2, x_3)$ to Bob.  Eve can observe only two of the three bits  sent by Alice
(i.e. $(x_1, x_2, \ast)$, $(x_1, \ast, x_3)$, or $(\ast, x_2, x_3)$), but not all of them. Alice can use this advantage Bob has
over Eve to send a one-bit secret message $m \in \{0, 1\}$ to
Bob such that Eve will consider both outcomes to be equally likely.  In
order to do this, Alice may make use of two fair coin tosses $(c_1, c_2)$, denoted as $0$ or $1$.
Then Alice chooses her channel inputs $(x_1, x_2, x_3)$ as follows:
	\begin{align*}
		(x_1, x_2, x_3) = (c_1, c_2, c_1 \oplus c_2 \oplus m)~,
	\end{align*}
where $\oplus$ is an XOR. Then Bob can decipher $m$ from his three channel inputs simply by XORing all his observations together. Eve, on the other hand, will have perfect equivocation on the value of $m$ regardless of which two bits she sees since all possible values are equiprobable regardless of the value of $m$.
\end{ex}

The next example highlights how secrecy can be attained in the source setting.

\begin{ex} \label{ex:intro_src}
(a) Consider the setting in which Alice is allowed to transmit one bit $x$ across a noiseless
public channel to Bob and Eve. Furthermore, Alice observes a two-bit string $(s_1, s_2)$ uniformly 
distributed over the set of all all 2-bit strings. Bob observes either the first bit $(s_1, \ast)$ or the second bit $(\ast, s_2)$ of Alice's string,
but not both, and Alice does not learn which of the two bits Bob observed. Eve observes nothing. Then, Alice and Bob can agree to make the secret key the first bit, and Alice's input to the channel can simply be the XOR of her two bits:
	\begin{align*}
		x = s_1 \oplus s_2~.
	\end{align*}
Then, Bob has enough information to determine the secret key, but Eve has perfect equivocation since she is equally likely to see $0$ or $1$ regardless of the value of the secret key.

(b) Suppose that Alice is allowed now to transmit two bits across the noiseless public 
channel $(x_1, x_2)$ to Bob and Eve, and the source observations are the same as in part (a). Instead of transmitting a secret key, Alice is given a secret message $m \in \{0, 1\}$ to communicate.
Then, Alice can simply transmit
	\begin{align*}
		(x_1, x_2) = (s_1 \oplus s_2, s_1 \oplus m)~,
	\end{align*}
which in effect uses the first channel symbol to construct a secret key as in part (a), and the second to use it as a one-time pad on the message. Since Bob can decode the secret key as earlier, Bob discovers the secret message. Eve, on the other hand, has perfect equivocation about $m$ since regardless of the message, all four values of $(x_1, x_2)$ are equiprobable.
\end{ex}

We now provide an example to illustrate how the above strategies can be combined.
\begin{ex} \label{ex:intro}
Suppose Alice has a three-bit noiseless
channel to Bob, and Eve can observe only two of the three bits as 
in Example \ref{ex:intro_ch}.
Additionally, Alice and Bob have source observations as in Example \ref{ex:intro_src}, where Eve observes no source.
The key idea is to combine the strategies used above, except to replace Alice's coin 
tosses $(c_1, c_2)$ in Example \ref{ex:intro_ch} with the input to the public channel 
from Example \ref{ex:intro_src}(b). Since Eve can learn the values of the coin tosses if she observes the first two channel inputs, under this strategy, those values function as a public bit pipe. This leads to the following channel inputs:
	\begin{align*}
		(c_1, c_2) &= (s_1 \oplus s_2, s_1 \oplus m_2) \\
		(x_1, x_2, x_3) &= (c_1, c_2, c_1 \oplus c_2 \oplus m_1) \\
					&= (s_1 \oplus s_2, s_1 \oplus m_2, s_2 \oplus m_1 \oplus m_2)
	\end{align*}
With this combined strategy, Alice can send a two-bit secret message $(m_1, m_2)$ to Bob, who can decode $m_1$ as in Example \ref{ex:intro_ch} and $m_2$ as in Example \ref{ex:intro_src}(b). Eve, on the other hand, has perfect equivocation about $(m_1,m_2)$ since regardless of their values, all possible values are equally likely  for whichever channel symbols she can observe.
\end{ex}

Example \ref{ex:intro} provides the essence of our separation approach for transmitting a secret message and matches the diagram shown in Figure \ref{fig:separation_strategy}:
	\begin{enumerate}
		\item Distill the channel into a public bit-pipe $(c_1, c_2)$ in addition to the private bit-pipe
		          $x_3$ over which the secret message is sent.
		\item Use part of the public bit-pipe to distill the sources and generate a secret key.
		\item Use the remainder of the public bit-pipe to send a secret message by using the secret 
		          key just generated as a one-time pad.
	\end{enumerate}
	
In the simple example above, we could exploit both the source and the
channel to the fullest as seen by comparing with Examples~\ref{ex:intro_ch}
and~\ref{ex:intro_src}(b). However, in general, it may be not desirable or even
possible for Alice to attempt to convey her source to Bob (for instance, if the
conditional entropy of the source at Alice conditioned on that at Bob is larger than the capacity of the channel). In the sequel, we will describe how the above strategy maps to this more general setting. The sketch of the strategy follows the spirit of Examples \ref{ex:intro_ch},
\ref{ex:intro_src}, and \ref{ex:intro} and as described above, provide an interpretation of the result as a separation strategy. 

\subsubsection{Case of no sources: Secrecy via the channel \label{subsec:channel}} 

Consider the case in which there is a noisy broadcast channel from Alice to
Bob and Eve; but there are no sources. Note that this resembles the cases studied in
Example \ref{ex:intro_ch} with the added wrinkle that the channel to Bob
may also be noisy. Recall that in Example \ref{ex:intro_ch}, given
sufficiently many fair coin tosses, Alice uses the channel to send a
message secretly to Bob. The work of Csisz\'{a}r and
K\"{o}rner~\cite{Csiszar78bcc} generalizes this approach as a means of
providing secrecy for all noisy broadcast channels. They also consider a
common message in addition to the private message and characterize the set
of all rate pairs such that the common message can be reliably recovered by
both Bob and Eve while the private message is recovered reliably by Bob,
but remains secret from Eve\footnote{In fact, they consider the
equivocation rate of Eve as a third parameter and characterize the rate
triple, but this is not relevant to our discussion.}. We may consider a
slight twist to this setting. We again consider two independent messages --
one {\em private} as in~\cite{Csiszar78bcc} and the other what we call {\em
public} -- both uniformly distributed over their alphabets both of which
need to be delivered reliably to Bob with the former remaining a secret
from Eve. The only difference from the setting of Csisz\'ar and K\"orner is
that we do not require the {public} message to be reliably recovered by
Eve. The following proposition can be proved directly following Csisz\'ar
and K\"orner (also see~\cite{xu2008bca}).
\begin{prop}\label{prop:secrecy_channel}For any given joint distribution of random variables
$V_1,V_2,X,Y,Z$ such that $V_2 - V_1 - X - (Y,Z)$ is a Markov chain
and the joint conditional distribution of $(Y,Z)$ given $X$ is consistent
with the given channel, the rate pair $(R_{\sf private},R_{\sf public})$ is achievable for the setting described above, where
\begin{align*}
R_{\sf private} &= [I(V_1;Y|V_2) - I(V_1;Z|V_2)]_+, \text{ and}\\
R_{\sf public} &= I(V_1;Y)-R_{\sf private}~.
\end{align*}
\end{prop}
\noindent The proof is a straightforward adaptation of the achievable strategy
in~\cite{Csiszar78bcc}\footnote{Roughly, the random coding argument runs as
follows: a $V_2$ codebook of rate $I(V_2;Y)$ is formed and a conditional
$V_1$ codebook of rate $I(V_1;Y)$ is formed for each $V_2$ codeword. The
conditional codebooks are binned so that the rate of each bin is
$I(V_1;Z|V_2)$. At Alice, a part of the public message bits worth rate
$I(V_2;Y)$ selects the $V_2$ codeword, the private message selects the
bin of the corresponding conditional $V_1$ codebook and the rest of the
public message selects the $V_1$ codeword within the bin. Bob performs
joint typical decoding. The reliability and secrecy of the scheme can be
shown along the lines of~\cite{Csiszar78bcc}.}. 

\subsubsection{Case of two noiseless bit pipes:  Private and
public \label{subsec:bitpipe}} 

Now consider the setting in which the channel is deterministic. In
particular, the channel is made up of two bit-pipes: (1) a {\em private}
bit-pipe of rate $R_{\sf private}$ which delivers its input bits from Alice
faithfully and only to Bob, and (2) a {\em public} bit-pipe of rate $R_{\sf
public}$ which delivers faithfully its input bits from Alice to both Bob and
Eve.

\begin{figure}[ht]
	\centering
        \begin{pspicture}(0,0.5)(6.7,4.55)

        \rput(-.5,3){\small{{Alice}}}%

        \rput(0,3){\psline{->}(0,0)(.5,0)}
        \rput(.2,3.2){\textcolor{black}{$M$}}

        \rput(0.5,3){\psframe(0,-0.4)(1.0,0.4)} %
        \rput(1,3){\small{\sc{Enc}}}%

        \rput(1,4.2){\psline{->}(0,0)(0,-0.8)}

        \rput(3.2,3.5){$R_{\sf private}$}
        \rput(3.2,3.2){\psline(-.1,-.1)(.1,.1)}
        \rput(1.5,3.2){\psline{->}(0,0)(3.4,0)}

        \rput(3.2,2.8){\psline(-.1,-.1)(.1,.1)}
        \rput(3.2,2.5){$R_{\sf public}$}
        \rput(1.5,2.8){\psline{->}(0,0)(3.4,0)}
        \rput(1.85,2.8){\psline(0,0)(0,-1.8)}
        \rput(1.85,1){\psline{->}(0,0)(3.05,0)}

        \rput(4.9,3){\psframe(0,-0.4)(1.0,0.4)} %
        \rput(5.4,3){\small{\sc{Dec}}}%

        \rput(3.2,4.2){sources}
        \rput(3.2,4.2){\psframe(-.75,-0.3)(0.75,0.3)} %

        \rput(5.4,4.2){\psline(0,0)(-1.45,0)}
        \rput(1,4.2){\psline(0,0)(1.45,0)}

        \rput(0.7,3.8){$S_A^n$}

        \rput(5.4,4.2){\psline{->}(0,0)(0,-0.8)}
        \rput(5.7,3.8){$S_B^n$}

        \rput(5.4,1){\small{{Eve}}}%

        \rput(5.9,3){\psline{->}(0,0)(.5,0)}
        \rput(6.2,3.25){\textcolor{black}{$\hat{M}$}}
        \rput(6.9,3){\small{{Bob}}}%

        \end{pspicture}
	\caption{Consider the case in which Alice and Bob share correlated source observations, 
	and there is both a private bit-pipe from Alice to Bob and a public bit-pipe from Alice to 
	Bob and Eve. This generalizes the problem considered in Example \ref{ex:intro_src}, and 
	the strategy considered in that setting generalizes naturally, as well.
	\label{fig:sec_separation}}
\end{figure}
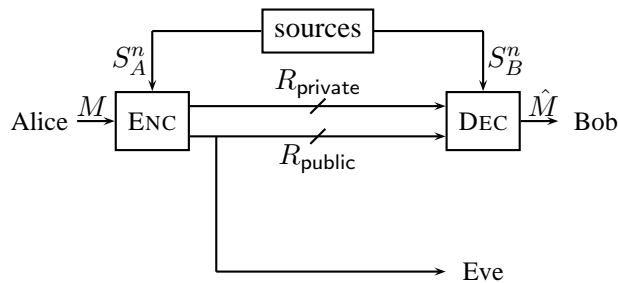

\paragraph{Secret-key only; no source observation at Eve} Consider the goal of generating the largest secret-key rate possible
when there is no source observation at Eve.\footnote{When Eve has a dependent source observation $S_E$, a further binning of
the codebook described in this section can be used to get a secret-key rate of
\[ R_{\sf SK} = I(U;S_B) - I(U;S_E) + R_{\sf private}, \]
where we restrict $U$ to those which satisfy
\[ I(U;S_A) - I(U;S_B) < R_{\sf public} + R_{\sf private},\]
and the Markov chain $U - S_A - (S_B,S_E)$.} This is reminiscent of Example \ref{ex:intro_src}(a) 
but with two added dimensions not present 
in that setting. First, there may not be 
enough rate on the public bit-pipe for Bob to determine Alice's source observation perfectly
to generate a secret key. A modified form of Wyner-Ziv's source coding strategy can be employed to handle this, which simply involves quantizing Alice's source and using that to generate the secret key.
Second, in addition to the public bit-pipe, 
there is also a private bit-pipe. Note that any component sent on the 
private bit-pipe is automatically a secret key, as well. Thus, if part of the bin index is sent on the private bit-pipe, it is also secret from Eve.

Then, given an auxiliary random variable $U$ which satisfies the
Markov chain $U - S_A - S_B$, a secret-key rate of $(I(U; S_B) + R) + R_{\sf private} - R$ is achievable if 
\begin{align*}
 I(U;S_A) - I(U;S_B) &\leq R_{\sf public} + R,\text{ and}\\
 R &\leq R_{\sf private}.
\end{align*}
Again, the work of 
Ahlswede-Csisz\'{a}r~\cite{AhlswedeCSecret93} can be used to show that this has the
required secrecy and uniformity properties, and thus, the resulting
secret-key rate is
\[ R_{\sf SK} = (I(U;S_B) + R) + (R_{\sf private} - R) = I(U;S_B) + R_{\sf
private}, \]
where we restrict $U$ to those which satisfy
\[ I(U;S_A) - I(U;S_B) < R_{\sf public} + R_{\sf private},\]
and the Markov chain $U - S_A - S_B$.

\paragraph{Secret message only; no source observation at Eve} Consider the case 
in which Alice desires to communicate a message
secretly at the largest possible rate when there is no source observation at Eve.\footnote{When Eve has a correlated source observation, a further binning of
the codebook described in this section can be used to get a secret message rate of
\[ R_{\sf SM} = R_{\sf private} + [I(U;S_B) - I(U;S_E)]_+,\]
where $U$ satisfies the Markov chain $U-S_A-(S_B,S_E)$ and the condition
\[  R_{\sf public} > I(U;S_A).\]} This scenario, depicted in Figure \ref{fig:sec_separation}, 
is a straightforward generalization of Example \ref{ex:intro_src}(b) from the introduction. In that 
example, Alice achieves secrecy across a public bit-pipe by binning her source observation 
based on Bob's side information to generate a shared secret key. On the rest of the public bit-pipe, Alice  uses this key as a one-time pad to send the secret message.

As earlier, there are two added dimensions in the current setting that are not present 
in Example \ref{ex:intro_src}(b). First, there may not be 
enough rate on the public bit-pipe for Bob to determine Alice's source observation perfectly
and thus generate a secret key. Again, Alice simply quantizes the source observation and applies 
the binning strategy as before, which corresponds to Wyner-Ziv's source coding scheme.
Second, in addition to the public bit-pipe, 
there is also a private bit-pipe. Because there is now a secret message, we split the message 
into two parts: the private bit-pipe is used fully to send part of the secret 
message (at rate $R_{\sf private}$), and the 
public bit-pipe is used as before to communicate the remaining bits secretly with the
correlated sources being exploited to provide the secrecy. 
However, since we now have 
to agree on \emph{specific} random bits instead of \emph{any} common random 
bits, we have two additional restrictions, which can cause the rate of the secret message to be lower than the secret key case above.
First, we have to reserve part of the public bit-pipe for sending the one-time padded secret message, which constrains part of the public bit-pipe rate $R_{\sf public}$ for generating the secret key from the sources. Second, sending part of the Wyner-Ziv bin index on the private bit-pipe will cost rate that 
can be used for sending a private message. Thus, it is better to reserve the private bit-pipe for sending a secret message, which costs  $R \leq R_{\sf private}$ bits that could have been used for generating the secret key, which means the rate of the secret key used as a one-time pad, and thus the effectiveness of 
the public bit-pipe, is significantly limited compared with the case of the secret key.

The work of Ahlswede-Csisz\'{a}r~\cite{AhlswedeCSecret93} can be
adapted to show that this approach satisfies the required secrecy and uniformity properties. The 
 secret-key is then used as a one-time pad to encrypt some extra messages
bits. Using this approach, given any joint distribution of 
$U - S_A - S_B$, a secret key of $I(U;S_B)$ can be generated by consuming 
$I(U;S_A) - I(U;S_B)$ bits from the public bit-pipe. This secret key can
then be used 
as a one-time pad on another $I(U;S_B)$ bits of the public bit-pipe   
to send a secret message of that rate. Hence, we must choose auxiliary 
random variable $U$ such that
\[ R_{\sf public} > \left(I(U;S_A) - I(U; S_B)\right) + I(U; S_B) =  I(U;S_A),\]
and the total secret message rate obtained is
\[ R_{\sf SM} = R_{\sf private} + I(U;S_B).\]
Unlike in the work of Csiszar-Narayan \cite{CsiszarNarayan2000}, in which
Alice and Bob only need to agree on {\em any} common random bits to
construct a secret key, for a secret message, we have the added constraint
that they must agree on {\em specific} random bits. Thus, the rates
achievable for secret message are less than those achievable for secret
key. 

\paragraph{Secret message -- secret-key tradeoff; no source observation at
Eve}

A secret-message -- secret-key tradeoff optimal strategy here\footnote{When Eve has a correlated source observation $S_E$, the tradeoff
becomes
\begin{align*}
R_{\sf SM} &\leq R_{\sf public} + R_{\sf private} - (I(U;S_A)-I(U;S_B)),
\text{ and}\\
R_{\sf SM} + R_{\sf SK} &\leq [I(U;S_B)-I(U;S_E)]_+ + R_{\sf private},
\end{align*}
where $U$ satisfies the Markov chain $U-S_A-S_B$ and the condition
\[ I(U;S_A)-I(U;S_B) \leq R_{\sf public} + R_{\sf private}.\]
} turns out to
be a natural combination of the above two: If (1) $R_{\sf SM}\leq R_{\sf
private}$, the secret-message is sent entirely over the private bit-pipe,
and the left-over rate ($R_{\sf private} - R_{\sf SM}$) of the private
bit-pipe rate along with the public bit-pipe is used for agreeing on a
secret-key from the correlated sources. This secret-key step is essentially
the {\em secret-key only} case discussed above. Otherwise, {\em i.e.}, if (2)
$R_{\sf SM} \geq R_{\sf private}$, all of the private bit-pipe is used to
carry a part of the secret message. For communicating the rest of the
secret message, at a rate of $R_{\sf SM} - R_{\sf private}$, and for
agreeing on a secret-key, the public bit-pipe and the sources are made use
of. The way the public bit-pipe is used is essentially the same as in the
{\em secret message only} case above. The only difference is that instead
of utilizing all of the secret-key generated from the sources as a one-time
pad to secure communication of a message over the public bit-pipe, here,
only a part of the secret-key is used for this purpose. The rate of the
unused part of the secret-key is $R_{\sf SK}$.

The resulting tradeoff is given by
\begin{align*}
R_{\sf SM} &\leq R_{\sf public} + R_{\sf private} - (I(U;S_A)-I(U;S_B)),
\text{ and}\\
R_{\sf SM} + R_{\sf SK} &\leq I(U;S_B) + R_{\sf private},
\end{align*}
where $U$ satisfies the Markov chain $U-S_A-S_B$ and the condition
\[ I(U;S_A)-I(U;S_B) \leq R_{\sf public} + R_{\sf private}.\]
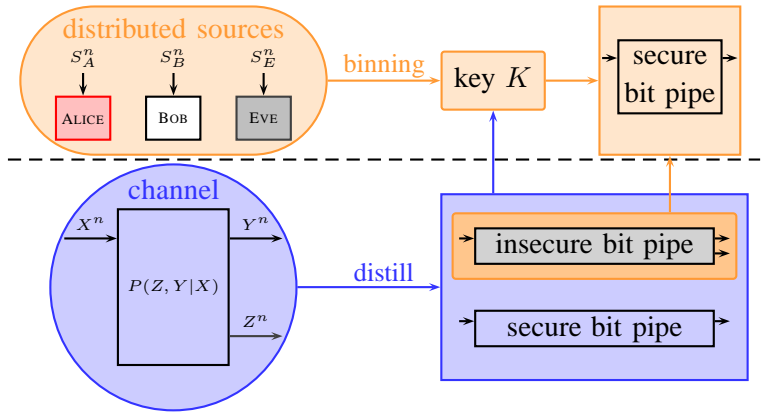
\begin{figure}[ht]
	\centering
        \begin{pspicture}(0,-0.5)(10,4.35)

        \rput(0,-0.5){\pscircle[linecolor=\darkblue, fillstyle=solid, fillcolor=\lightblue](2.2,1){1.65}}
        \rput(2.2,1.8){\textcolor{\darkblue}{channel}}

        \rput(5,0.7){\small{\textcolor{\darkblue}{distill}}}
        \rput(3.85,0.5){\psline[linecolor=\darkblue]{->}(0,0)(1.9,0)}
        \rput(7.8,0.5){\psframe[linecolor=\darkblue, fillstyle=solid, fillcolor=\lightblue](-2.05,1.25)(2.05,-1.25)}

        \rput(0,2.2){\psline[linestyle=dashed](0,0)(10,0)}
        \rput(2.2,2.75){\psframe[linecolor=\darkorange, fillstyle=solid, fillcolor=\lightorange, framearc=1](-2.05,1.5)(2.05,-.5)}
        \rput(2.2,3.95){\textcolor{\darkorange}{distributed sources}}

        \rput(5,3.45){\small{\textcolor{\darkorange}{binning}}}
        \rput(4.25,3.25){\psline[linecolor=\darkorange]{->}(0,0)(1.5,0)}

        \rput(0,2.75){
                \rput(0.5,0){\psframe[linecolor=red, fillstyle=solid, fillcolor=pink](0.125,-0.3)(.875,0.3)} %
                \rput(1,0){\tiny{\sc{Alice}}}%

                \rput(1,0.6){\psline{->}(0,0)(0,-0.3)}
                \rput(1,0.8){\tiny{$S_A^n$}}

                \rput(1.7,0){\psframe[linecolor=black, fillstyle=solid, fillcolor=white](0.125,-0.3)(.875,0.3)} %
                \rput(2.2,0){\tiny{\sc{Bob}}}%

                \rput(2.2,0.6){\psline{->}(0,0)(0,-0.3)}
                \rput(2.2,0.8){\tiny{$S_B^n$}}

                \rput(2.9,0){\psframe[linecolor=gray!50!black, fillstyle=solid, fillcolor=gray!50!white](0.125,-0.3)(.875,0.3)} %
                \rput(3.4,0){\tiny{\sc{Eve}}}%

                \rput(3.4,0.6){\psline{->}(0,0)(0,-0.3)}
                \rput(3.4,0.8){\tiny{$S_E^n$}}
        }
        \rput(0,-0.5){
                \rput(1.1,1.85){\tiny{$X^n$}}
                \rput(0.75,1.65){\psline{->}(0,0)(.7,0)}

                \rput(1.2,1){\psframe(0.25,-1.05)(1.75,1.05)} %
                \rput(2.2,1){\tiny{$P(Z,Y|X)$}}%

                \rput(3.3,1.85){\tiny{$Y^n$}}
                \rput(2.95,1.65){\psline[linecolor=black]{->}(0,0)(0.7,0)}

                \rput(3.3,0.55){\tiny{$Z^n$}}
                \rput(2.95,0.35){\psline[linecolor=gray!50!black]{->}(0,0)(0.7,0)}
        }

        \rput(5.35,3.25){\rput(0,0){\psframe[linecolor=\darkorange, fillstyle=solid, fillcolor=\lightorange, framearc=.1](0.4,-.4)(1.8,0.4)}
        \rput(1.1,0){key $K$}}
        \rput(7.15,3.25){\psline[linecolor=\darkorange]{->}(0,0)(0.7,0)}
        \rput(7.8,1.05){\psframe[linecolor=\darkorange, fillstyle=solid, fillcolor=\medorange, framearc=.1](-1.9,-.45)(1.9,.45)}

        \rput(6.0,1.15){\psline{->}(0,0)(.2,0)}
        \rput(9.4,1.15){\psline{->}(0,0)(.2,0)}
        \rput(9.4,0.95){\psline{->}(0,0)(.2,0)}

        \rput(7.8,1.05){\psframe[fillstyle=solid, fillcolor=gray!35!white](-1.6,.25)(1.6,-.25)}
        \rput(7.8,1.05){\textcolor{black}{insecure bit pipe}}

        \rput(6.0,.05){\psline{->}(0,0)(.2,0)}
        \rput(9.4,.05){\psline{->}(0,0)(.2,0)}
        \rput(7.8,-.05){\psframe[fillstyle=solid, fillcolor=\lightblue](-1.6,.25)(1.6,-.25)}
        \rput(7.8,-.05){\textcolor{black}{secure bit pipe}}

        \rput(6.45,1.75){\psline[linecolor=\darkblue]{->}(0,0)(0,1.1)}
        \rput(8.8,1.5){\psline[linecolor=\darkorange]{->}(0,0)(0,0.75)}

        \rput(9.8,3.25){\psframe[linecolor=\darkorange, fillstyle=solid, fillcolor=\lightorange](-1.95,1)(-.05,-1)}
        \rput(8.8,3.3){\psframe[fillstyle=solid, fillcolor=\lightorange](-.7,.5)(.7,-.5)}
        \rput(7.9,3.55){\psline{->}(0,0)(.2,0)}
        \rput(9.5,3.55){\psline{->}(0,0)(.2,0)}
        \rput(8.8,3.55){\textcolor{black}{secure}}
        \rput(8.8,3.05){\textcolor{black}{bit pipe}}

        \end{pspicture}
	\caption{The intuition behind Theorem~\ref{thm:achievability}. In
this approach, the channel is distilled into a public bit pipe and a private one.
The sources take advantage of part of the rate from each of these channels to
generate a secret key. This key is divided into the final secret key and a
one-time pad, the latter of which is used to secure the remainder of the public bit pipe
for sending part of the secret message. The remainder of the private bit
pipe is used to send the remainder of the secret message.
\label{fig:separation_strategy}}
\end{figure}

\subsubsection{The General case}

Now let us turn to the general case with sources in which the channel is
not necessarily deterministic. This resembles Example \ref{ex:intro}, and
as in that case, we can apply a combination of the strategies in Section
\ref{subsec:channel} and \ref{subsec:bitpipe}. Indeed, by treating the
random coin tosses Alice uses in Proposition \ref{prop:secrecy_channel} as
a public bit-pipe, we can construct a public and private bit-pipe from the
channel and can leverage the source strategy from Section
\ref{subsec:bitpipe}. This approach enables us to obtain the rates in
\eqref{eq:inner1} and \eqref{eq:inner2}. However, we should note that
neither the independence requirement nor the uniformity requirement in
Proposition \ref{prop:secrecy_channel} hold for the messages sent over the
bitpipes in~\ref{subsec:bitpipe}, though they may hold approximately.  And
hence, this discussion does not constitute a proof of
Theorem~\ref{thm:achievability}. Formalizing the above is an alternative
approach to proving Theorem~\ref{thm:achievability}, but we do not pursue
it here. A schematic interpretation of the discussion in this section is
shown in Figure \ref{fig:separation_strategy}.

\section{Discussion \label{sect:secrecy_discussion}}

\subsection{Extensions and Additional Results}

\paragraph{Stochastically degraded sources and channels}
There are several additional results related to the present work that we wish to note. 
For instance, it turns out that the result presented 
in Theorem \ref{thm:outerbounds} holds more generally than the degradedness 
conditions outlined. First, the degradedness conditions can be relaxed to
stochastically degraded conditions for both the source and channels. This
simply involves a slightly more cumbersome argument in our converse proof,
but no changes to the achievable strategy are necessary. For completeness,
the converse argument is given in Appendix~\ref{app:outerbounds}.

\paragraph{Bandwidth mismatch}
We only considered the case of matched bandwidths, i.e., there is one
source symbol per every channel symbol. Our results can be readily extended
when there is a bandwidth mismatch of say $m_S$ source symbols for every
$m_C$ channel symbols. By considering a vector source with $m_S$ symbols
and vector channel with $m_C$ symbols, we can directly invoke
Theorem~\ref{thm:achievability}. Further, by restricting the auxiliary
random variables to be i.i.d. across the vector components we can arrive at
the following achievable region. Let ${\mathcal P}_\text{\sf mismatch}$ be the
set of all joint distributions $p$ of random variables
$U_1,V_1,V_2,X,Y,Z,S_A,S_B,S_E$ which satisfy the same conditions as in the
definition of ${\mathcal P}$ except for \eqref{eq:pcondition} being
replaced by
\[ m_C I(V_1;Y) \geq m_S (I(U_1;S_A) - I(U_1;S_B).\]
Let ${\mathcal R}_\text{\sf mismatch}(p)$, for $p\in{\mathcal P}_\text{\sf
mismatch}$ be the set of all rate pairs $(R_{\sf SK},R_{\sf SM})$ which
satisfy
\begin{align*}
R_{\sf SM} &\leq I(V_1;Y) - \frac{m_S}{m_C}( I(U_1 ; S_A) -  I(U_1;S_B) ),\\
R_{\sf SK}  + R_{\sf SM} &\leq [I(V_1;Y|V_2) - I(V_1;Z|V_2)]_+ 
         + \frac{m_S}{m_C}[I(U_1;S_B) - I(U_1;S_E)]_+.
\end{align*} 
Then the set ${\mathcal C}_{\sf mismatch}$ of all achievable rate pairs,
where rates are measured per channel use, satisfies
\[ {\mathcal C}_{\sf mismatch} \supseteq \bigcup_{p\in{\mathcal P}_{\sf
mismatch}} {\mathcal R}_{\sf mismatch}(p).\] 

For the degraded case considered in Theorem~\ref{thm:outerbounds}, we can
also show the optimality of the above achievable region under bandwidth
mismatch. Appendix~\ref{app:outerbounds} discusses the modifications needed
in the converse. A consequence of this is that the optimality of Gaussian
signalling shown in Proposition~\ref{prop:Gaussianexample} continues to
hold even under bandwidth mismatch. 

\paragraph{Strong secrecy}
All the secrecy results in this paper can be directly strengthened by
dropping the ${1/n}$ factor in \eqref{eq:secrecy_cond} without any penalty on
the rates achieved. This follows directly from the work of Maurer and
Wolf~\cite{MaurerWolf00} on privacy amplification using extractors. Maurer
and Wolf demonstrate this for the problems of secret key agreement of
Ahlswede and Csisz\'ar, and secure message transmission of Csisz\'ar and
K\"orner. The key idea is to perform several (independent) repetitions of
the scheme which produces weakly secure keys and achieves weakly secure
data transmission. A privacy amplification step using an extractor can be
employed on the weakly secure keys to generate a strongly secure key. The
privacy amplification step involves Alice using a small (polylogarithmic in
blocklength) purely random key which she needs to share with Bob over a
public channel. Alice will need to use the broadcast channel to do this,
but the overhead involved is negligible and does not affect the rates
achieved. To send a strongly secure message in addition to generating a
strongly secure key, Alice will first invert the extractor operation to
produce the equivalent weakly secure messages that when passed through the
extractor would produce the strongly secure message she intends to
transmit. Then she proceeds to transmit these equivalent weakly secure
messages using the scheme in this paper. The small key is also sent
separately using a channel code. At the end of all the transmissions, Bob
who will have recovered all the weakly secure keys and weakly secure
messages as well as the small key can invoke the extractor to recover the
strongly secure key and the strongly secure message. 

\paragraph{Two extensions of Theorem~\ref{thm:outerbounds}}
Two other extensions of the results in Theorem \ref{thm:outerbounds} were
shown in \cite{PrabhakaranRamchandran07}. First, given only the sources and
a public bit-pipe from Alice to Bob and Eve, the condition under which
Alice and Bob cannot generate a positive rate secret-key is in fact weaker
than the case where the sources are degraded in favor of Eve\footnote{This
condition which can be inferred from \cite{AhlswedeCSecret93} is that for
every $\tilde{U}_1,\tilde{U}_2$ satisfying the Markov chain $\tilde{U}_2 -
\tilde{U}_1 - S_A - (S_B,S_E)$,
\[I(\tilde{U}_1;S_B|\tilde{U}_2)\leq I(\tilde{U}_1;S_E|\tilde{U}_2).\]}.
Under this weaker condition, it was shown
in~\cite{PrabhakaranRamchandran07} that the optimal strategy involves
ignoring the sources, and utilizing only the channel. In particular,
${\mathcal R}(p)$ is now the set of all non-negative rate pairs satisfying
the condition
\[R_{\sf SK}+R_{\sf SM} = [I(V_1;Y)-I(V_1;Z)]_+,\]
where $V_1 - X - (Y,Z)$ is a Markov chain.
Thus the optimal strategy in this case reduces to that of Csisz\'{a}r and
K\"{o}rner~\cite{Csiszar78bcc}, and there is essentially no distinction
between sending a secret message and generating a secret-key.  

Second, a channel degraded in favor of Eve is a condition
under which the channel resource by itself cannot provide any secrecy, but
note that the condition under which the channel resource cannot provide any
secrecy is looser than this type of degradation. This condition is when the
channel to Eve is `less noisy' than the channel to Bob~\cite[Corollary~3,
pg.~341]{Csiszar78bcc}.  Under this looser condition, but when the
source component degraded in favor of Eve is absent,
the optimality of turning the channel into a public bit-pipe was shown
in~\cite{PrabhakaranRamchandran07} for secret-key generation. In the
special case where Eve has no source observation, this optimality was shown
for secret communication as well.

\paragraph{Secure source-channel coding}
Note that sending a secret message is equivalent to the case in which Alice must 
send a discrete uniform source losslessly to Bob that must be kept secret from 
Eve. A straightforward extension of our result for the secret message case, 
shown in \cite{EswaranRamchandran:Allerton08}, demonstrates that optimality continues to 
hold if one is interested in reconstructing any discrete memoryless source, both for the 
lossless and lossy cases. In this situation, an additional layer of separation between 
the private bit pipes and the compression of the source can be shown to establish the result.

\subsection{Open problems}
The above extensions do not close the door on this problem, and there are several 
considerations that currently warrant further research. Indeed, the general rate 
region and structure of optimal strategies are still open problems. One avenue is to 
consider extensions of the result beyond the degraded case and beyond some of the extensions 
discussed above.

Another interesting avenue to consider is the setting in which the sources and channel are correlated. Note that 
in such a setting, there may not be a clean distinction between a source observation and a channel 
output at either Bob or Eve, which resembles the setup for Theorem \ref{thm:joint_encoding}. Indeed, the strategy and proof presented for Theorem \ref{thm:joint_encoding} continues to hold if the sources and channels are correlated.

Furthermore, the setting of the strategy presented in Theorem \ref{thm:joint_encoding} 
coincides with a problem studied by Chen and Vinck \cite{ChenVinck}, in which Alice must send Bob a secret message (i.e., $R_{\sf SK}=0$), Alice has non-causal state information about the channel. Chen and Vinck make the 
additional assumption that Eve observes degraded versions of Bob's channel outputs, but 
Theorem \ref{thm:joint_encoding} holds even without this assumption. In fact, when Chen and Vinck's scheme is considered in the context of Theorem~\ref{thm:outerbounds} (i.e., independent sources and channels), but with the degradedness condition of Chen and Han Vinck (i.e., there is no reversely degraded channel component), we already know from Theorem~\ref{thm:outerbounds} that the secrecy capacity is given by 
\begin{align}
C_\text{\sf SM} &= \max \min\{I(X_F;Y_F) - I(U;S_{A}|S_{B}),
         I(X_F;Y_F|V) - I(X_F;Z_F|V) + I(U;S_{B}|S_{E})\},
\label{eq:degradedSep}
\end{align}
where the maximization is over joint distributions of the form
$p_{V,X_F}p_{U,S_A}$. Chen and Han Vinck's achievable secrecy rate is
\begin{align}
R &= \max \min\{I(W;Y)-I(W;S_A),I(W;Y)-I(W;Z)\}\nonumber\\
  &= \max \min\{I(W;Y_F,S_B)-I(W;S_A), I(W;Y_F,S_B)-I(W;Z_F,S_E)\},
\label{eq:Vinck}
\end{align}
where the maximization is over $p_{W,X_F|S_A}$. Whenever the maximizer
of \eqref{eq:degradedSep} is such that $V$ is a constant, we may
choose $W=(X_F,U)$ in \eqref{eq:Vinck} to match the capacity. For instance,
in the Gaussian example of Section~\ref{sec:gaussian_example}, it is indeed the
case that optimal joint distribution involves a constant $V$. Note that the Gaussian
case of Chen and Vinck's scheme was first considered by
Mitrpant, Vinck and Luo~\cite{MitrpantVinckLuo06}. But in
general, it does not appear to be the case that \eqref{eq:Vinck} equals the
secrecy capacity in \eqref{eq:degradedSep}.

While Chen and Vinck present an upper bound on the secret message rate, 
it does not coincide with either their achievable strategy or Theorem 
\ref{thm:joint_encoding}. The work in \cite{eswaranpr:asilomar08} provides a marginal improvement to 
the upper bound presented by Chen and Vinck, but the problems of characterizing the 
rate region and optimal strategies remain open. Indeed, this region may also be 
tightened by improving upon the achievable strategy in Theorem \ref{thm:joint_encoding}.

Progress on any of these fronts could lead to new insights on how strategies 
may optimally combine source and channel resources for secrecy, as well as on 
the interplay between secret keys and messages.

\begin{appendices}

\section{Proof of Theorem \ref{thm:joint_encoding}}
\label{app:achievabilityproof}

Since a secret message automatically satisfies the constraints of a secret key, 
it is enough to prove that the following $(R_{\sf SK},R_{\sf SM})$ pair is
achievable.
\begin{align*}
R_{\sf SM} &= \min( I(\mathbf{U}; \mathbf{Y}) - I(\mathbf{U}; {S}),
   I(\mathbf{U}; \mathbf{Y} | \mathbf{V}) 
       - I(\mathbf{U}; \mathbf{Z} | \mathbf{V}) ), \text{ and}\\
R_{\sf SK} &= [ I(\mathbf{U}; \mathbf{Y} | \mathbf{V}) 
       - I(\mathbf{U}; \mathbf{Z} | \mathbf{V}) 
        - ( I(\mathbf{U}; \mathbf{Y}) - I(\mathbf{U}; {S}) ) ]_+\\
  &= [ I(\mathbf{U}; {S}) - I(\mathbf{V}; \mathbf{Y}) 
           - I(\mathbf{U}; \mathbf{Z} | \mathbf{V}) ]_+.
\end{align*}

We divide the proof into two cases. In each case, we use a random coding argument to 
show the existence of a codebook for which the probability of an encoding error at Alice,
decoding error at Bob, and decoding error at Eve given additional side information are 
all small. We then show that such a code satisfies the secrecy and uniformity conditions. \\
\subsection{Case 1: $I(\mathbf{U};S) \leq I(\mathbf{V};  \mathbf{Y}) +
I(\mathbf{U}; \mathbf{Z} |  \mathbf{V})$}
In this case, we need only prove that the pair
\begin{align*}
R_{\sf SM} &= I(\mathbf{U}; \mathbf{Y} | \mathbf{V}) 
       - I(\mathbf{U}; \mathbf{Z} | \mathbf{V}), \text{ and}\\
R_{\sf SK} &= 0
\end{align*}
is achievable.

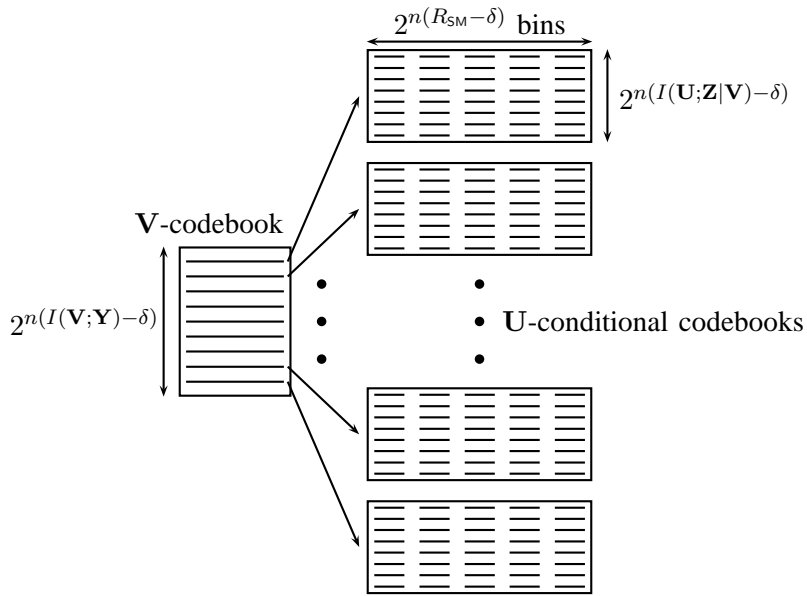
\begin{figure}[ht]
	\centering
        \begin{pspicture}(-3,-4)(5.75,4)

        \rput(0,0){\psframe(-1.5,-1)(0,1)}

        \rput(-1.1,1.3){$\bf V$-codebook}

        \multirput{0}(-1.4,-.8)(0,.2){9}{
                \psline(0,0)(1.3,0)
        }

        \psline{<->}(4.2,2.385)(4.2,3.625)
        \rput(5.5,3){$2^{n(I(\mathbf{U};\mathbf{Z}|\mathbf{V})-\delta)}$}

        \psline{<->}(-1.7,-1)(-1.7,1)
        \rput(-2.75,0){$2^{n(I(\mathbf{V};\mathbf{Y})-\delta)}$}

        \psline{->}(-.05,.8)(.9,3)
        \psline{->}(-.05,.6)(.9,1.5)

        \psline{->}(-.05,-.6)(.9,-1.5)
        \psline{->}(-.05,-.8)(.9,-3)

        \psdots(.4,0)
        \psdots(.4,-.5)
        \psdots(.4,.5)

        \rput(2.5,4){$2^{n(R_{\sf SM} - \delta)}$ bins}
        \rput(4.8,0){$\mathbf{U}$-conditional codebooks}

        \psline{<->}(1,3.725)(4,3.725)

        \multirput{0}(1,-3)(0,1.5){2}{
                \psframe(0,-0.625)(3,0.625)

                \multirput{0}(.1,-.525)(0,.15){8}{
                        \multirput{0}(0,0)(0.6,0){5}{
                                \psline(0,0)(0.4,0)
                        }
                }
        }

        \psdots(2.5,0)
        \psdots(2.5,-.5)
        \psdots(2.5,.5)

        \multirput{0}(1,3)(0,-1.5){2}{
                \psframe(0,-0.625)(3,0.625)

                \multirput{0}(.1,-.525)(0,.15){8}{
                        \multirput{0}(0,0)(0.6,0){5}{
                                \psline(0,0)(0.4,0)
                        }
                }
        }

        \end{pspicture}
	\caption{The codebook used for Case 1 of the achievable strategy consists of 
	a $\mathbf{V}$-codebook, each codeword of which indexes a conditional codebook.
	The bins in the conditional codebook correspond directly to the private bit-pipe, and 
	the $\mathbf{V}$-codebook and codewords in each bin to the public bit-pipe. Analogously, the codewords in each conditional codebook correspond to quantization points for the source $S^n$.
	\label{fig:secrecy_case1_bitpipes}}
\end{figure}

\noindent{\em Random Coding Argument}

\paragraph{Codebook generation} We create a codebook of blocklength $n$ with
$2^{n(I(\mathbf{U};\mathbf{Y}) - 3\delta)}$ elements composed of two parts.  We
create a blocklength-$n$ $\mathbf{V}$-codebook of size
$2^{n(I(\mathbf{V};\mathbf{Y}) - \delta)}$ by drawing the codewords uniformly
from the $\epsilon$-strongly typical set~\cite[Chapter 1.2]{CsiszarKornerbook1ed}
$\typical_\epsilon^{\ast(n)}$ of $n$-length $\mathbf{V}$ sequences. Let us index these
codewords using $i\in \{ 1, \ldots, 2^{n(I(\mathbf{V};\mathbf{Y}) -
\delta)}\}$. For each such codeword $\mathbf{v}^n(i)$, a conditional $\mathbf{U}$-codebook of size
$2^{n\left(I(\mathbf{U};\mathbf{Y} | \mathbf{V}) - 2\delta\right)}$ is created
by drawing the codewords uniformly from the set of all $n$-length $\mathbf{U}$~sequences
which are conditionally $\epsilon$-strongly typical conditioned on the
$\mathbf{V}$~sequence $\mathbf{v}^n$. For each such conditional codebook, we distribute these
sequences into $2^{n(R_{\sf SM}
- \delta)}$ {\em bins} such that each bin contains $2^{n(I(\mathbf{U};
  \mathbf{Z} | \mathbf{V})-\delta)}$ codewords, indexing each bin by $m \in
\{1,\ldots, 2^{n(R_{\sf SM} - \delta)}\}$. Let the codewords in each bin be
indexed by $j \in \{ 1, \ldots, 2^{n(I(\mathbf{U}; \mathbf{Z} | \mathbf{V}) -
\delta)}\}$. 

Note that 
there is a direct correspondence between the bins, and the private bit-pipe, with 
the codewords in each bin and the $\mathbf{V}$-codebook corresponding to the 
public bit-pipe of the separation strategy. Furthermore, as will be seen in encoding, the $\bf U$-codewords 
are simply quantization points for the source 
$S^n$. A schematic of this codebook is depicted in 
Figure \ref{fig:secrecy_case1_bitpipes}.

In this separation context, Case 1 refers to the scenario in which there is 
insufficient randomness from the source $S^n$ alone to determine 
the input to the public bit-pipe. Thus, we further divide the set of all $\mathbf{U}$-codewords into
$2^{n(I(\mathbf{V};\mathbf{Y}) + I(\mathbf{U};\mathbf{Z}|\mathbf{V}) -
I(\mathbf{U};{S})-3\delta)}$ {\em buckets}\footnote{For there to be at least one bucket, we require that $3\delta < I(\mathbf{V};\mathbf{Y}) + I(\mathbf{U};\mathbf{Z}|\mathbf{V}) -
I(\mathbf{U};{S})$. However, this is not an issue since we will take $\delta \to 0$.}  as follows: if (i) $I(\mathbf{U};S)
\geq I(\mathbf{V};\mathbf{Y})$, that is, 
there is sufficient randomness in the source to determine the $\mathbf{V}$-codeword completely,
 we divide up the codewords
in each bin of every conditional codebook among the buckets such that each
bucket has the same number of codewords. Thus, in each bin of each of the
conditional codebooks there are
\begin{align*}
2^{n(I(\mathbf{U}; \mathbf{Z} | \mathbf{V}) - \delta 
- I(\mathbf{V};\mathbf{Y}) - I(\mathbf{U};\mathbf{Z}|\mathbf{V})
+ I(\mathbf{U};{S}) + 3\delta)}
= 2^{n(I(\mathbf{U};{S}) - I(\mathbf{V};\mathbf{Y}) + 2\delta)}
\end{align*}
codewords which belong to a given bucket. If (ii) $I(\mathbf{U}; S) <
I(\mathbf{V};\mathbf{Y})$, then, the $\mathbf{U}$-codewords are divided up among the
buckets such that every bucket has no more than one codeword
which belongs to the same bin of a conditional codebook. In this case, for
a given bucket, there are 
\begin{align*}
2^{n(I(\mathbf{U};\mathbf{Y}) - 3\delta -
(R_{\sf SM} - \delta) - (I(\mathbf{V};\mathbf{Y}) +
I(\mathbf{U};\mathbf{Z}|\mathbf{V}) - I(\mathbf{U};{S}) - 3\delta))}
= 2^{n( I(\mathbf{U};{S}) + \delta)}
\end{align*}
codewords each belonging to a different conditional codebook and holding
the same bin index. The buckets are indexed by $k\in\{ 1, \ldots,
2^{n(I(\mathbf{V};\mathbf{Y}) + I(\mathbf{U};\mathbf{Z}|\mathbf{V}) -
I(\mathbf{U};S)-3\delta)}\}$. For a $\mathbf{U}$-codeword, we will explicity
indicate its bucket index along with the conditional codebook it belongs
to, its bin-index and its index within the bin as $\mathbf{u}^n(i,m,j,k)$. 

\paragraph{Encoding} Let $m \in \{1, \ldots, 2^{n(R_{\sf SM} - \delta)}\}$
index the secret message.  To send $m$, using her private random string
$\Phi_{A}$, Alice obtains a $\Phi_{\sf bucket}$ which is uniformly
distributed over the set $\{1, \ldots,$ $2^{n(I(\mathbf{V}; \mathbf{Y}) +
I(\mathbf{U}; \mathbf{Z} | \mathbf{V}) - I(\mathbf{U};{S})-3\delta)}\}$,
assigns $k=\Phi_{\sf bucket}$, and looks in bin $m$ (of all the conditional
codebooks) for a $\mathbf{V}^n(i),\mathbf{U}^n(i,m,j,k)$ such that
$(\mathbf{V}^n(i),\mathbf{U}^n(i,m,j,k), S^n)$ are jointly typical. Thus,
the $\mathbf{U}$ codeword is selected such that it belongs to bin $m$ and bucket
$k=\Phi_{\sf bucket}$ and such that it is jointly typical with the source
observation ${S}^n$. If more than one choice is found, Alice chooses one of
them arbitrarily. If none are found, Alice declares an error. A test
channel $p_{X|\mathbf{U}, S}$ stochastically generates the channel input
$X^n$.

The probability of encoding failure can be bounded as follows. In case~(i), 
\begin{align*}
P_e\leq& {\sf Pr}(S^n \notin \typical_\epsilon^{\ast(n)}) + \sum_{s^n\in \typical_\epsilon^{\ast(n)}} 
    p_S(s^n) \sum_k p_{\Phi_{\sf bucket}}(k) \cdot \\
    &\left\{\left[
       \sum_{\mathbf{v}^n \in  \typical_\epsilon^{\ast(n)}} 
          {\sf Pr}(\mathbf{V}^n=\mathbf{v}^n)
          \left[
             1 - {\sf Pr}((\mathbf{v}^n,\mathbf{U}^n, s^n)\in
                 \typical_\epsilon^{\ast(n)}|\mathbf{V}^n=v^n)
          \right]^{2^{n(I(\mathbf{U};\mathbf{S}) - I(\mathbf{V};\mathbf{Y})
 + 2\delta)}}
    \right]^{2^{n(I(\mathbf{V};\mathbf{Y})-\delta)}}\right\}.
\end{align*}
where the term ${\sf Pr}(\mathbf{V}^n=\mathbf{v}^n)$ is evaluated with the distribution
for $\mathbf{V}^n$ being given by the uniform distribution over all
$\epsilon$-strongly typical $\mathbf{v}^n$ sequences, and ${\sf
Pr}((v^n,\mathbf{U}^n, s^n)\in \typical_\epsilon^{\ast(n)}|\mathbf{V}^n=v^n)$ is
evaluated with the distribution for $\mathbf{U}^n$ being given by the
uniform distribution over all $\mathbf{u}^n$ sequences which are
conditionally $\epsilon$-strongly typical with $\mathbf{v}^n$. Since $\mathbf{V}$
and $S$ are independent, for $s^n\in \typical_\epsilon^{\ast(n)}$ and
$\mathbf{v}^n \in \typical_\epsilon^{\ast(n)}$, this probability is
\begin{align}
{\sf Pr}((\mathbf{v}^n,\mathbf{U}^n, s^n)\in \typical_\epsilon^{\ast(n)}|\mathbf{V}^n=\mathbf{v}^n)
 \geq 2^{-n(I(\mathbf{U};S|\mathbf{V}) + \epsilon_1)}. \label{eq:probofencodingerrorcomp1}
\end{align}
Here $\epsilon_1 \to 0$ as $\epsilon \to 0$. This will also be the case for 
any future subscripted $\epsilon_{\#}$ in the sequel.
Using this in the term within the braces in the upperbound for $P_e$ 
and simplifying
\begin{align*}
&\left[
       \sum_{\mathbf{v}^n \in  \typical_\epsilon^{\ast(n)}} 
          {\sf Pr}(\mathbf{v}^n)
          \left[
             1 - {\sf Pr}((\mathbf{v}^n,\mathbf{U}^n, s^n)\in
                 \typical_\epsilon^{\ast(n)}|\mathbf{V}^n=v^n)
          \right]^{2^{n(I(\mathbf{U};{S}) - I(\mathbf{V};\mathbf{Y})
+2\delta)}}
    \right]^{2^{n(I(\mathbf{V};\mathbf{Y})-\delta)}}\\
&\leq \left[ \left[ 1 - 2^{-n(I(\mathbf{U};S|\mathbf{V}) +
\epsilon_1)}\right]^{2^{n(I(\mathbf{U};{S}) - I(\mathbf{V};\mathbf{Y})
 +2\delta)}}
    \right]^{2^{n(I(\mathbf{V};\mathbf{Y})-\delta)}}\\
&\stackrel{\text{(a)}}{\leq} 
      e^{-2^{n(I(\mathbf{U};{S})+\delta)}
          2^{-n(I(\mathbf{U};{S}|\mathbf{V})+\epsilon_1)}}\\
&\stackrel{\text{(b)}}{=} e^{-2^{n(\delta - \epsilon_1)}},
\end{align*}
where (a) follows from $(1-x)^n\leq e^{-nx}$, and (b) from the fact that 
$I(\mathbf{U};{S}) = I(\mathbf{V},\mathbf{U};{S}) = I(\mathbf{U};{S}|\mathbf{V})$ which in turn is
a consequence of the Markov chain $\mathbf{V} - \mathbf{U} - S$ and the independence of $\mathbf{V}$ and $S$.
Substituting this in the upperbound for $P_e$,
\begin{align*}
P_e\leq& {\sf Pr}(S^n \notin \typical_\epsilon^{\ast(n)}) + \left(\sum_{s^n\in \typical_\epsilon^{\ast(n)}} 
    p_S(s^n) \sum_k p_{\Phi_{\sf bucket}}(k)\right) \cdot e^{-2^{n(\delta - \epsilon_1)}}\\
   =& {\sf Pr}(S^n \notin \typical_\epsilon^{\ast(n)}) + 
      (1-{\sf Pr}(S^n \notin \typical_\epsilon^{\ast(n)}))e^{-2^{n(\delta - \epsilon_1)}}.
\end{align*}
Thus, we can make $P_e$ as small as desired by choosing sufficiently small $\delta$, $\epsilon$ ($\delta >
\epsilon_1$), and sufficiently large $n$.

Under case~(ii), the probability of encoding failure can be similarly
bounded. Now, we have
\begin{align*}
P_e\leq&  {\sf Pr}(S^n \notin \typical_\epsilon^{\ast(n)}) + \sum_{s^n\in \typical_\epsilon^{\ast(n)}} 
    p_S(s^n) \sum_k p_{\Phi_{\sf bucket}}(k) \cdot\\
    &\left\{\left[
       \sum_{\mathbf{v}^n \in  \typical_\epsilon^{\ast(n)}} 
          {\sf Pr}(\mathbf{v}^n)
          \left[
             1 - {\sf Pr}((\mathbf{v}^n,\mathbf{U}^n, s^n)\in
                 \typical_\epsilon^{\ast(n)}|\mathbf{V}^n=v^n)
          \right]\right]^{2^{n(I(\mathbf{U};\mathbf{S})+\delta)}}\right\}.
\end{align*}
where the term ${\sf Pr}((v^n,\mathbf{U}^n, s^n)\in
\typical_\epsilon^{\ast(n)}|\mathbf{V}^n=v^n)$ can be evaluated as in
\eqref{eq:probofencodingerrorcomp1}. Substituting this above and following
similar steps, by choosing sufficiently small $\delta$, $\epsilon$ ($\delta
> \epsilon_1$), and sufficiently large $n$, we can make $P_e$ as small as
desired.

\paragraph{Decoding at Bob} Bob receives $\mathbf{Y}^n$ and searches for a
unique $(\mathbf{V}^n, \mathbf{U}^n)$ pair such that $(\mathbf{V}^n,
\mathbf{U}^n, \mathbf{Y}^n)$ that are $\epsilon$-strongly jointly typical.
If no such pair exists, Bob declares an error. Otherwise, Bob identifies
the corresponding bin-index $\hat{m}$, and declares this the secret
message. Hence, conditioned on encoding being successful, a decoding error
results only if there is a $\hat{m}\neq m$ such that, there are $\hat{i}$,
$\hat{j}$, $\hat{k}$ such that $(\mathbf{v}^n(\hat{i}),
\mathbf{u}^n(\hat{i},\hat{m},\hat{j},\hat{k}), \mathbf{Y}^n)$ are
$\epsilon$-strongly jointly typical. Using the union bound, we can
upperbound the probability of this by
\begin{align*}
\sum_{\hat{i}} \sum_{\hat{m}\neq m} \sum_{\hat{j}}
  &{\sf Pr}\left((\mathbf{V}^n(\hat{i}), \mathbf{U}^n(\hat{i},\hat{m},\hat{j}),
            \mathbf{Y}^n)\in \typical_\epsilon^{\ast(n)}\right)\\
&= \sum_{\hat{i}\neq i} \sum_{\hat{m}\neq m} \sum_{\hat{j}}
  {\sf Pr}\left((\mathbf{V}^n(\hat{i}), \mathbf{U}^n(\hat{i},\hat{m},\hat{j}),
            \mathbf{Y}^n)\in \typical_\epsilon^{\ast(n)}\right)\\
 &\quad + \sum_{\hat{m}\neq m} \sum_{\hat{j}}
  {\sf Pr}\left((\mathbf{V}^n(i), \mathbf{U}^n(i,\hat{m},\hat{j}),
            \mathbf{Y}^n)\in \typical_\epsilon^{\ast(n)}\right)\\
&\leq 2^{n(I(\mathbf{U};\mathbf{Y}) - 3\delta)}
       2^{-n(I(\mathbf{U};\mathbf{Y}) - \epsilon_2)} + 
     2^{n(R_{\sf SM}-\delta)}
      2^{n(I(\mathbf{U}; \mathbf{Z} | \mathbf{V})-\delta)}
       2^{-n(I(\mathbf{U}; \mathbf{Y} | \mathbf{V}) - \epsilon_3)}
\end{align*}
which can be made as small as desired by choosing sufficiently small
$\delta$, $\epsilon$ ($\delta > \epsilon_2, \epsilon_3$), and sufficiently large $n$.
       
\paragraph{Decoding at Eve with side information} Consider Eve who has
access to $M, \mathbf{V}^n$. Then, the bin in which a potential
$\mathbf{U}^n$ exists is known to be at most $2^{n(I(\mathbf{U};\mathbf{Z}
| \mathbf{V}) - \delta)}$. We may upperbound the probability of decoding
error as we did above.  Consider the jointly typical decoder for
$\mathbf{U}^n$ given $\mathbf{Z}^n$ in this bin. There are two error
events: $E_1$ is the event no sequence in the bin is jointly typical with
$\mathbf{Z}^n$, and $E_2$ is the event a false sequence in the subbin is
jointly typical with $\mathbf{Z}^n$. We have, ${\sf Pr}(E_1) \to 0$ as $n
\to \infty$ and the probability a false sequence is jointly typical with
$\mathbf{Z}^n$ is $2^{-n(I(\mathbf{U};\mathbf{Z} | \mathbf{V}) -
\epsilon_4)}$. By a union bound, we can make the probability of error as
small as desired by choosing sufficiently small $\delta$, $\epsilon$
($\delta > \epsilon_4$), and sufficiently large $n$.

By the usual random coding arguments, we may now conclude that for any
$\delta>0$, for sufficiently large $n$, there exists a codebook (with rates as
in the codebook construction above) such that (i) Bob can recover the secret
message with probability of error not larger than $\delta$ and (ii) Eve, when
provided with the message and the $\mathbf{V}$ codeword, can recover the
$\mathbf{U}$ codeword with probability of error not larger than $\delta$. We
now simply have to verify that this codebook also has the property that Eve's
information about the message (given $\mathbf{Z}^n$) is small, {\em i.e.}, the secrecy
condition.

\noindent{\em Proof of Secrecy Condition}.\\
First observe that
\begin{align}
	H(M | \mathbf{Z}^n) &\geq H(M | \mathbf{Z}^n, \mathbf{V}^n) \notag\\
	&= H(M, \mathbf{Z}^n| \mathbf{V}^n)  - H(\mathbf{Z}^n|
\mathbf{V}^n) \notag\\
	&= H(M, \mathbf{U}^n, \mathbf{Z}^n| \mathbf{V}^n) - H(\mathbf{U}^n
| M, \mathbf{Z}^n, \mathbf{V}^n) - H(\mathbf{Z}^n| \mathbf{V}^n) \notag\\
	&\stackrel{\text{(a)}}{\geq} H(\mathbf{U}^n, \mathbf{Z}^n|
\mathbf{V}^n) - H(\mathbf{U}^n | M, \mathbf{Z}^n, \mathbf{V}^n) -
H(\mathbf{Z}^n| \mathbf{V}^n) \notag\\
	&= H(\mathbf{U}^n| \mathbf{V}^n) + H(\mathbf{Z}^n|\mathbf{U}^n,
\mathbf{V}^n) - H(\mathbf{U}^n| M, \mathbf{Z}^n, \mathbf{V}^n) -
H(\mathbf{Z}^n| \mathbf{V}^n). \label{eq:secrecycondition1}
\end{align}
where (a) follows from non-negativity of conditional entropy. We now bound
each of these terms.

Let us define, for every $\mathbf{u}^n(i,m,j,k)$ codeword 
\begin{align*}
 E = \{ s^n : \exists (i,m,j,k) \text{ such that }
 (\mathbf{v}^n(i), \mathbf{u}^n(i,m,j,k),s^n) \in \typical_\epsilon^{\ast (n)}\}
\end{align*}
Recall that for all $\alpha > 0$, there exists $n$ sufficiently large
such that decoding (and hence encoding) succeeds with probability greater
than $1 - \alpha$, {\em i.e.,}
	\begin{align*}
		\PP (S^n \in E) \geq 1 -  \alpha~.
	\end{align*}
Furthermore, the probability 
	\begin{align*}
		\PP (&(\mathbf{V}^n,\mathbf{U}^n)=(\mathbf{v}^n(i),\mathbf{u}^n(i,m,j,k)), S^n \in E)
 \notag \\
		&= \PP ( M=m, \Phi_{\sf bucket} = k, (\mathbf{V}^n,
\mathbf{U}^n) = (\mathbf{v}^n(i),\mathbf{u}^n(i,m,j,k)), S^n \in E) \notag \\
		&\leq \PP(M = m) \cdot 2^{ - n(I(\mathbf{V}; \mathbf{Y}) +
I(\mathbf{U}; \mathbf{Z} | \mathbf{V}) - I(\mathbf{U}; S)-3\delta) } \cdot
\sum_{s^n: (\mathbf{v}^n(i),\mathbf{u}^n(i,m,j,k),s^n) \in \typical_\epsilon^{\ast
(n)}} \PP(S^n = s^n) \\
		&\leq 2^{ -n (R_{\sf SM} - \delta)} \cdot 2^{ - n(I(\mathbf{V}; \mathbf{Y}) +
I(\mathbf{U}; \mathbf{Z} | \mathbf{V}) - I(\mathbf{U}; S)-3\delta) } \cdot  2^{ n
H(S | \mathbf{U}) + n \epsilon} \cdot 2^{ - n H(S) + n \epsilon } \\
		&= 2^{-n(R_{\sf SM} - \delta)} \cdot 2^{ - n(I(\mathbf{V}; \mathbf{Y}) +
I(\mathbf{U}; \mathbf{Z} | \mathbf{V}) - I(\mathbf{U}; S)-3\delta) } \cdot  2^{ - n
I(S ; \mathbf{U}) + 2 n \epsilon } \\
		&= 2^{-n(R_{\sf SM} - \delta)} \cdot 2^{ - n(I(\mathbf{V}; \mathbf{Y}) +
I(\mathbf{U}; \mathbf{Z} | \mathbf{V})-3\delta) + 2 n \epsilon } ~,
\intertext{which along with the lowerbound on $\PP(S^n \in E)$ above implies
that}
\PP(&(\mathbf{V}^n,\mathbf{U}^n)=(\mathbf{v}^n(i),\mathbf{u}^n(i,m,j,k))|S^n
\in E)\\
 &\leq 2^{-nR_{\sf SM}} \cdot 2^{ - n(I(\mathbf{V}; \mathbf{Y}) +
I(\mathbf{U}; \mathbf{Z} | \mathbf{V})) + n \epsilon_5}\\
       &= 2^{-nI(\mathbf{U};\mathbf{Y}) + n\epsilon_5}.
	\end{align*}
Also, we know that the size of
the codebook in which $(\mathbf{V}^n,\mathbf{U}^n)$ take values is
less than $2^{nI(\mathbf{U};\mathbf{Y})}$ which implies that
\begin{align*}
H(\mathbf{U}^n,\mathbf{V}^n|S^n\in E) \geq nI(\mathbf{U};\mathbf{Y})
  - n\epsilon_5.
\end{align*}
Using this we can bound the first term in \eqref{eq:secrecycondition1}.
	\begin{align*} 
		H(\mathbf{U}^n| \mathbf{V}^n) &= H(\mathbf{U}^n, \mathbf{V}^n) - H(\mathbf{V}^n) \\
		&\stackrel{(a)}{\geq} H(\mathbf{U}^n, \mathbf{V}^n) - n(I(\mathbf{V};\mathbf{Y})) \\
		&\stackrel{(b)}{\geq} H(\mathbf{U}^n, \mathbf{V}^n | S^n \in E) \cdot \PP(S^n \in E) - n(I(\mathbf{V};\mathbf{Y})) \\
		&= nI(\mathbf{U};\mathbf{Y}) - nI(\mathbf{V};\mathbf{Y}) -
n\epsilon_6\\
                &= nI(\mathbf{U};\mathbf{Y}|\mathbf{V}) - n \epsilon_6,
	\end{align*}
where (a) follows from the fact that $\mathbf{V}^n$ takes values in a codebook whose
size is smaller than $2^{nI(\mathbf{V};\mathbf{Y})}$, and (b) follows from the fact that
conditioning reduces entropy.

We bound the second term in \eqref{eq:secrecycondition1} as follows
\begin{align*}
H(\mathbf{Z}^n|\mathbf{U}^n,\mathbf{V}^n) 
  &= H(\mathbf{Z}^n|\mathbf{U}^n)\\
  &= \sum_{\mathbf{u}^n} Pr(\mathbf{U}^n=\mathbf{u}^n)
H(\mathbf{Z}^n|\mathbf{U}^n=\mathbf{u}^n)\\
  &\stackrel{(a)}{=}  \sum_{\mathbf{u}^n} Pr(\mathbf{U}^n=\mathbf{u}^n)
         \sum_{\mu \in {\mathcal U}} N(\mu|\mathbf{u}^n)
H(\mathbf{Z}|\mathbf{U}=\mu)\\
  &\stackrel{(b)}{\geq} \sum_{\mathbf{u}^n} Pr(\mathbf{U}^n=\mathbf{u}^n)
         \sum_{\mu \in {\mathcal U}} n(Pr(U=\mu) - \epsilon)
H(\mathbf{Z}|\mathbf{U}=\mu)\\
  &= \sum_{\mathbf{u}^n} Pr(\mathbf{U}^n=\mathbf{u}^n)
         (n H(\mathbf{Z}|\mathbf{U}) - n\epsilon_7)\\
  &= nH(\mathbf{Z}|\mathbf{U}) - n\epsilon_7,
\end{align*}
where (a) follows from the memoryless nature of the virtual channel from
$\mathbf{U}$ to $\mathbf{Z}$ and $N(\mu|\mathbf{u}^n)$ counts the number of times $\mu$
appears in the codeword $\mathbf{u}^n$, and (b) follows from the fact that
all the $\mathbf{u}^n$ codewords belong to $\typical_\epsilon^{\ast (n)}$.
Note that from (a) onwards, we use ${\mathbf U},{\mathbf Z}$ to denote a
pair of random variables distributed according to the joint distribution
$p_{{\mathbf U},{\mathbf Z}}$.

The third term can be bounded by using Fano's inequality and the fact that
Eve can recover the $\mathbf{U}^n$ codeword with a probability of error
$\epsilon$ when she has access to $M$ and $\mathbf{V}^n$ in addition to her
observation $\mathbf{Z}^n$.
\begin{align*}
	H(\mathbf{U}^n | M, K, \mathbf{Z}^n, \mathbf{V}^n) \leq 1 + n \cdot
\epsilon \cdot I(\mathbf{U};\mathbf{Z} | \mathbf{V}) = n\epsilon_8.
\end{align*}

Finally, to bound the fourth term, let $T$ be an indicator random variable
which takes on the value 1 when $(\mathbf{V}^n,\mathbf{Z}^n) \in
\typical_\epsilon^{\ast (n)}$ and 0 otherwise.
\begin{align}
H(\mathbf{Z}^n|\mathbf{V}^n) 
 &\leq H(\mathbf{Z}^n, T |\mathbf{V}^n) \notag\\
 &\leq 1 + H(\mathbf{Z}^n|\mathbf{V}^n, T=1)Pr(T=1) +
   n\log|{\mathcal Z}|Pr(T=0). \label{eq:secrecycondition1term4}
\end{align}
But
\[ Pr(T=0)=Pr((\mathbf{V}^n,\mathbf{Z}^n) \notin \typical_\epsilon^{\ast (n)}) \leq
\epsilon_9.\]
Furthermore, we have
\begin{align*}
H(\mathbf{Z}^n|\mathbf{V}^n, T=1) 
  &= \sum_{\mathbf{v}^n} Pr(\mathbf{V}^n=\mathbf{v}^n|T=1) 
                         H(\mathbf{Z}^n|\mathbf{V}^n=\mathbf{v}^n, T=1)\\
  &\stackrel{(a)}{\leq} \sum_{\mathbf{v}^n} Pr(\mathbf{V}^n=\mathbf{v}^n|T=1)
                      \log |\typical_\epsilon^{\ast(n)}(p_{\mathbf{Z}|\mathbf{V}}|\mathbf{v}^n)|\\
  &\leq \sum_{\mathbf{v}^n} Pr(\mathbf{V}^n=\mathbf{v}^n|T=1)
                      (nH(\mathbf{Z}|\mathbf{V}) + n\epsilon)\\
  &= nH(\mathbf{Z}|\mathbf{V}) + n\epsilon,
\end{align*}
where in (a) we used $|\typical_\epsilon^{\ast(n)}(p_{\mathbf{Z}|\mathbf{V}}|\mathbf{v}^n)|$ to
denote the size of the set of all $\mathbf{z}^n$ such that
$(\mathbf{z}^n,\mathbf{v}^n)\in \typical_\epsilon^{\ast(n)}$. Thus,
\eqref{eq:secrecycondition1term4} becomes
\begin{align*}
H(\mathbf{Z}^n|\mathbf{V}^n) \leq nH(\mathbf{Z}|\mathbf{V}) + n \epsilon_{10}.
\end{align*}

Hence, we may conclude from \eqref{eq:secrecycondition1} that
\begin{align*}
\frac{1}{n}H(M|\mathbf{Z}^n) &\geq I(\mathbf{U};\mathbf{Y}|\mathbf{V}) +
H(\mathbf{Z}|\mathbf{U}) - H(\mathbf{Z}|\mathbf{V}) + \epsilon_{11}\\
  &= I(\mathbf{U};\mathbf{Y}|\mathbf{V}) -
I(\mathbf{U};\mathbf{Z}|\mathbf{V}) + \epsilon_{11}\\
  &= R_{\sf SM} + \epsilon_{11}.
\end{align*}
Thus we have shown the secrecy condition.

\vspace{1cm}
\subsection{Case 2: $I(\mathbf{U};S) > I(\mathbf{V};  \mathbf{Y}) +
I(\mathbf{U}; \mathbf{Z} |  \mathbf{V})$}
In this case, we only need to show the achievability of
\begin{align*}
R_{\sf SM} &= I(\mathbf{U}; \mathbf{Y}) - I(\mathbf{U}; \mathbf{S}),
   \text{ and}\\
R_{\sf SK} &= I(\mathbf{U}; \mathbf{S}) - I(\mathbf{V}; \mathbf{Y}) 
           - I(\mathbf{U}; \mathbf{Z} | \mathbf{V}).
\end{align*}
We proceed as in case~1. Note that below we assume $R_{\sf SM}>0$. If $R_{SM}=0$, the only modification needed is to avoid the binning step associated with the secret message.\\
\noindent{\em Random Coding Argument}

\paragraph{Codebook Generation} We generate a codebook of blocklength-$n$ with
$2^{n(I(\mathbf{U};\mathbf{Y}) - 2\delta)}$ elements composed of two parts.
The first part is a blocklength $n$ $\mathbf{V}$-codebook of size
$2^{n(I(\mathbf{V};\mathbf{Y}) - \delta)}$ codewords by drawing the
codewords uniformly from the $\epsilon$-strongly typical set
$\typical_\epsilon^{\ast(n)}$ of $n$-length $\mathbf{V}$ sequences, indexing each
by $i \in \{1, \ldots, 2^{n(I(\mathbf{V};\mathbf{Y}) - \delta)}\}$. For
each codeword $\mathbf{v}^n$, a conditional $\mathbf{U}$-codebook of size
$2^{n\left(I(\mathbf{U};\mathbf{Y} | \mathbf{V}) - \delta\right)}$ is
created by drawing the codewords uniformly from the set of $n$-length
$\mathbf{U}$~sequences which are conditionally $\epsilon$-strongly typical
conditioned on the $\mathbf{V}$~sequence $\mathbf{v}^n$.  For each
conditional codebook, we distribute these sequences among $2^{n(R_{\sf
SM}-3\delta)}$ {\em bins} such that each bin contains $2^{n
(I(\mathbf{U};S) - I(\mathbf{V}; \mathbf{Y})+2\delta)}$ codewords. We index
the bins by $m$, where $m \in \{1,\ldots, 2^{n (R_{\sf SM} - 3\delta)}\}$.
The sequences in each bin are assigned to $2^{n (R_{\sf SK}+3\delta)}$ {\em
subbins} so that each subbin contains $2^{n (I(\mathbf{U};\mathbf{Z} |
\mathbf{V}) - \delta)}$ codewords, indexing each subbin by $k \in \{1,
\ldots, 2^{n(R_{\sf SK}+3\delta)} \}$.
We index each of the elements in the subbin by $\ell \in \{ 1, \ldots, 2^{n
(I(\mathbf{U};\mathbf{Z} | \mathbf{V}) - \delta)} \}$, and denote the
specific index as $\Phi_{\sf sub-index}$. For a $\mathbf{U}$-codeword, we
will explicitly indicate its index as $\mathbf{u}^n(i,m,k, \ell)$. 

\paragraph{Encoding}
Let $m \in \{1, \ldots, 2^{n(R_{\sf SM}-3\delta)}\}$ index the secret
message.  For this fixed $m$, Alice selects a $\mathbf{V}^n (i),
\mathbf{U}^n(i,m,k, \ell)$ such that $(\mathbf{V}^n,\mathbf{U}^n(i,m,k,
\ell), S^n)$ are jointly typical. If none are found, Alice declares an
error. A test channel $p_{X|\mathbf{U}, S}$ stochastically encodes the
channel input $X^n$. The subbin index $k$ is set as the secret~key. Note that the
secret~key is determined automatically by the $\mathbf{U}^n(i,m,k, \ell)$
selected.

For a fixed $m$, and the probability of an encoding failure is given by
\begin{align*}
P_e\leq& {\sf Pr}(S^n\notin \typical_\epsilon^{\ast(n)}) + \sum_{s^n\in \typical_\epsilon^{\ast(n)}} 
    p_S(s^n) \cdot\\
    &\left\{\left[
       \sum_{\mathbf{v}^n \in  \typical_\epsilon^{\ast(n)}} 
          {\sf Pr}(\mathbf{V}^n=\mathbf{v}^n) 
          \left[
             1 - {\sf Pr}((\mathbf{v}^n,\mathbf{U}^n, s^n)\in
                 \typical_\epsilon^{\ast(n)}|\mathbf{V}^n=\mathbf{v}^n)
          \right]^{2^{n(I(\mathbf{U};S) - I(\mathbf{V}; \mathbf{Y})+2\delta)}}
    \right]^{2^{n(I(\mathbf{V};\mathbf{Y})-\delta)}}\right\}.
\end{align*}
where the term ${\sf Pr}(\mathbf{V}^n=\mathbf{v}^n)$ is evaluated with the
distribution for $\mathbf{V}^n$ being given by the uniform distribution
over all $\epsilon$-strongly typical $\mathbf{v}^n$ sequences, ${\sf
Pr}((v^n,\mathbf{U}^n, s^n)\in \typical_\epsilon^{\ast(n)}|\mathbf{V}^n=v^n)$ is
evaluated with $\mathbf{U}^n$ being uniformly distributed over the set
$\typical_\epsilon^{\ast(n)}(p_{\mathbf{U}|\mathbf{V}}|\mathbf{V}^n=v^n)$ of all
$\mathbf{u}^n$ sequences which are conditionally $\epsilon$-strongly
typical with $\mathbf{v}^n$. As in case~1, since $\mathbf{V}$ and $S$ are
independent, for $s^n\in \typical_\epsilon^{\ast(n)}$ and $\mathbf{v}^n \in
\typical_\epsilon^{\ast(n)}$,
\begin{align*}
{\sf Pr}((\mathbf{v}^n,\mathbf{U}^n, s^n)\in \typical_\epsilon^{\ast(n)}|\mathbf{V}^n=\mathbf{v}^n)
 &\geq 2^{-n(I(\mathbf{U};S|\mathbf{V}) + \epsilon_1)} \\
 &= 2^{-n(I(\mathbf{U};S) + \epsilon_1)}.
\end{align*}
Using this in the term within the braces in the upperbound for $P_e$ 
and simplifying
\begin{align*}
	&\left[
       \sum_{\mathbf{v}^n \in  \typical_\epsilon^{\ast(n)}} 
          |\typical_\epsilon^{\ast(n)}(p_\mathbf{V})|^{-1} \cdot
          \left[
             1 - {\sf Pr}((\mathbf{v}^n,\mathbf{U}^n, s^n)\in
                 \typical_\epsilon^{\ast(n)}|\mathbf{V}^n=v^n)
          \right]^{2^{n(I(\mathbf{U};S) - I(\mathbf{V}; \mathbf{Y})+2\delta)}}
    \right]^{2^{n(I(\mathbf{V};\mathbf{Y})-\delta)}} \\
    &\leq \left[
       \sum_{\mathbf{v}^n \in  \typical_\epsilon^{\ast(n)}} 
          |\typical_\epsilon^{\ast(n)}(p_\mathbf{V})|^{-1} \cdot
          \left[
             1 - 2^{-n(I(\mathbf{U};S) + \epsilon_1)}
          \right]^{2^{n(I(\mathbf{U};S) - I(\mathbf{V}; \mathbf{Y})+2\delta)}}
    \right]^{2^{n(I(\mathbf{V};\mathbf{Y})-\delta)}} \\
    &= \left[
          \left[
             1 - 2^{-n(I(\mathbf{U};S|\mathbf{V}) + \epsilon_1)}
          \right]^{2^{n(I(\mathbf{U};S) - I(\mathbf{V}; \mathbf{Y})+2\delta)}}
    \right]^{2^{n(I(\mathbf{V};\mathbf{Y})-\delta)}} \\
    &\stackrel{(a)}{\leq}
          e^{- 2^{-n(I(\mathbf{U};S) + \epsilon_1)} \cdot
2^{n(I(\mathbf{U};S)+\delta)}},
\end{align*}
where (a) follows from the inequality $1 -x  \leq e^{-x}$.
Substituting this in the upperbound for $P_e$, as in case~1, we can make $P_e$ as small
as desired by choosing sufficiently small $\delta$, $\epsilon$ ($\delta >
\epsilon_1$), and sufficiently large $n$.

\paragraph{Decoding at Bob}
Bob receives $\mathbf{Y}^n$ and searches for a unique
$(\mathbf{V}^n(\hat{i}), \mathbf{U}^n(\hat{i},\hat{m},\hat{k},\hat{\ell}))$
pair such that $(\mathbf{V}^n(\hat{i}), 
\mathbf{U}^n(\hat{i},\hat{m},\hat{k},\hat{\ell}), \mathbf{Y}^n)\in
\typical_\epsilon^{\ast(n)}$. If no such pair exists, Bob declares an error.
Otherwise Bob declares $\hat{m}$ to be the secret message and $\hat{k}$ to
be the secret key. Conditioned on encoding being successful,
an error results only if there is a pair
$(\hat{m},\hat{k}) \neq (m,k)$ such that there are $\hat{i}$ and $\hat{\ell}$
and $\left(\mathbf{V}^n(\hat{i}), \mathbf{U}^n(\hat{i},\hat{m},\hat{k}_1,
\hat{\ell}), \mathbf{Y}^n \right) \in \typical_\epsilon^{\ast(n)}$. We can upperbound
the probability of this by
\begin{align*}
\sum_{\hat{i}} \sum_{(\hat{m},\hat{k}) \neq (m,k)} \sum_{\hat{\ell}}
  &{\sf Pr}\left((\mathbf{V}^n(\hat{i}),
           \mathbf{U}^n(\hat{i},\hat{m},\hat{k},\hat{\ell}),
            \mathbf{Y}^n)\in \typical_\epsilon^{\ast(n)}\right)\\
&= \sum_{\hat{i}\neq i} \sum_{(\hat{m},\hat{k}) \neq (m,k)} \sum_{\hat{\ell}}
  {\sf Pr}\left((\mathbf{V}^n(\hat{i}), 
           \mathbf{U}^n(\hat{i},\hat{m},\hat{k},\hat{\ell}),
            \mathbf{Y}^n)\in \typical_\epsilon^{\ast(n)}\right)\\
 &\quad +\sum_{(\hat{m},\hat{k}) \neq (m,k)} \sum_{\hat{\ell}}
  {\sf Pr}\left((\mathbf{V}^n(i), \mathbf{U}^n(i,\hat{m},\hat{k},\hat{\ell}),
            \mathbf{Y}^n)\in \typical_\epsilon^{\ast(n)}\right)\\
&\leq 2^{n(I(\mathbf{U};\mathbf{Y}) - 2\delta)}
       2^{-n(I(\mathbf{U};\mathbf{Y}) - \epsilon_2)} + 
      2^{n(I(\mathbf{U}; \mathbf{Y} | \mathbf{V})-\delta)}
       2^{-n(I(\mathbf{U}; \mathbf{Y} | \mathbf{V}) - \epsilon_3)}~,
\end{align*}
which can be made as small as desired by choosing sufficiently small $\delta$,
$\epsilon$ ($\delta > \epsilon_2,\epsilon_3$), and sufficiently large $n$.

\paragraph{Decoding at Eve with side information} Consider Eve who has
access to $M, K, \mathbf{V}^n$. Then, the subbin in which a potential
$\mathbf{U}^n$ exists is known to be at most $2^{n(I(\mathbf{U};\mathbf{Z}
| \mathbf{V}) - \delta)}$.  The probability of error of the jointly typical
decoder for $\mathbf{U}^n$ in this bin given $\mathbf{Z}^n$ can be bounded
as above. There are two error events: $E_1$ is the event no sequence in the
bin is jointly typical with $\mathbf{Z}^n$, and $E_2$ is the event a false
sequence in the subbin is jointly typical with $\mathbf{Z}^n$. We have,
$P(E_1) \to 0$ as $n \to \infty$ and the probability that a
false sequence is jointly typical with $\mathbf{Z}^n$ is
$2^{-n(I(\mathbf{U};\mathbf{Z} | \mathbf{V}) - \epsilon_4)}$.
By a union bound, we can make the probability of error as
small as desired by choosing sufficiently small $\delta$, $\epsilon$
($\delta > \epsilon_4$), and sufficiently large $n$.

By the usual random coding arguments, as in case~1, we may now conclude that
for any $\delta>0$, for sufficiently large $n$, there exists a codebook with
the rates as set above, such that (i) Bob can recover the secret message and
the secret key with the probability of error not larger than $\delta$ and (ii)
Eve, when provided with the message and the $\mathbf{V}$ codeword, can recover
the $\mathbf{U}$ codeword with probability of error not larger than $\delta$.
We now have to verify that this implies that (1) Eve's information about the
message (given $\mathbf{Z}^n$) goes to zero (secrecy condition) and (2) the
secret key is approximately uniformly distributed over its alphabet (uniformity
condition).

\noindent{\em Proof of Secrecy Condition}.\\
First we observe that 
\begin{align}
	H(M, K | \mathbf{Z}^n) &\geq H(M, K | \mathbf{Z}^n, \mathbf{V}^n) \notag \\
	&= H(M, K, \mathbf{Z}^n | \mathbf{V}^n)  - H(\mathbf{Z}^n | \mathbf{V}^n) \notag \\
	&= H(M, K, \mathbf{U}^n, \mathbf{Z}^n | \mathbf{V}^n) - H(\mathbf{U}^n | M, K, \mathbf{Z}^n, \mathbf{V}^n) - H(\mathbf{Z}^n | \mathbf{V}^n) \notag\\
	&\stackrel{(a)}{\geq} H(\mathbf{U}^n , \mathbf{Z}^n | \mathbf{V}^n) - H(\mathbf{U}^n | M, K, \mathbf{Z}^n,   \mathbf{V}^n) - H(\mathbf{Z}^n | \mathbf{V}^n) \notag \\
	&= H(\mathbf{U}^n | \mathbf{V}^n) + H(\mathbf{Z}^n | \mathbf{U}^n, \mathbf{V}^n) - H(\mathbf{U}^n | M, K, \mathbf{Z}^n,   \mathbf{V}^n) - H(\mathbf{Z}^n | \mathbf{V}^n)~, \label{eq:secrecy_condition1_case2}
\end{align}
where (a) follows from $H(M, K | \mathbf{U}^n, \mathbf{Z}^n, \mathbf{V}^n)  \geq 0$. We now bound
each of these terms.
Let us define, for every $\mathbf{u}^n(i,m,k, \ell)$ codeword 
\begin{align}
 E = \{ s^n : \exists (i,m,k, \ell) \text{ such that }
 (\mathbf{v}^n(i), \mathbf{u}^n(i,m,k, \ell),s^n) \in \typical_\epsilon^{\ast (n)}\}
\end{align}
Recall that for all $\alpha > 0$, there exists $n$ sufficiently large
such that decoding (and hence encoding) succeeds with probability greater
than $1 - \alpha$, {\em i.e.,}
	\begin{align*}
		\PP (S^n \in E) \geq 1 -  \alpha~.
	\end{align*}
Furthermore, the probability 
	\begin{align*}
		\PP &\left((\mathbf{V}^n, \mathbf{U}^n) = (\mathbf{v}^n, \mathbf{u}^n(i,m,k, \ell)), S^n \in E \right) \notag \\
		&= \PP \left(M = m, K= k, (\mathbf{V}^n, \mathbf{U}^n) = (\mathbf{v}^n, \mathbf{u}^n(i,m,k, \ell)), S^n \in E \right) \\
		&= \PP \left(M = m\right) \cdot
 \sum_{s^n:(\mathbf{v}^n(i),\mathbf{u}^n(i,m,j,k),s^n) \in \typical_\epsilon^{\ast(n)}} \PP\left(S^n = s^n\right)\\
		&\leq 2^{-n (I(\mathbf{U};\mathbf{Y}) - I(\mathbf{U}; S) - 3\delta)} \cdot 2^{nH(S | \mathbf{U}) + n\epsilon} 2^{-n H(S) + n\epsilon} \\
		&= 2^{-n I(\mathbf{U};\mathbf{Y}) + 3n \delta + 2n \epsilon}~, 
	\end{align*}
which, along with the lower bound on $\PP (S^n \in E)$ above implies that 
	\begin{align}
		\PP &\left((\mathbf{V}^n, \mathbf{U}^n) = (\mathbf{v}^n, \mathbf{u}^n(i,m,k, \ell))| S^n \in E \right) \notag \\
		&\leq 2^{-n I(\mathbf{U};\mathbf{Y}) + n \epsilon_{12}}~. \label{eq:prob_unif_bound}
	\end{align}
Also, we know that the size of the codebook in which 
$(\mathbf{V}^n, \mathbf{U}^n)$ take values is less than $2^{n I(\mathbf{U}; \mathbf{Y})}$, 
which implies that 
	\begin{align*}
		H(\mathbf{U}^n, \mathbf{V}^n | S^n \in E) \geq n I(\mathbf{U};\mathbf{Y}) - n \epsilon_{12} ~.
	\end{align*}
Using this, we can bound the first term in \eqref{eq:secrecy_condition1_case2}:
	\begin{align*}
		H(\mathbf{U}^n | \mathbf{V}^n) &= H(\mathbf{U}^n, \mathbf{V}^n) - H(\mathbf{V}^n) \\
		&\stackrel{(a)}{\geq} H(\mathbf{U}^n, \mathbf{V}^n) - n (I(\mathbf{V}; \mathbf{Y})) \\
		&\stackrel{(b)}{\geq}  H(\mathbf{U}^n, \mathbf{V}^n | S^n \in E) \cdot \PP (S^n \in E) 
		- n I(\mathbf{V}; \mathbf{Y}) \\
		&= n I(\mathbf{U};\mathbf{Y}) - n I(\mathbf{V}; \mathbf{Y}) - n \epsilon_{13}~,
	\end{align*}
where (a) follows from the fact that $\mathbf{V}^n$ takes values in a codebook whose size is 
smaller than $2^{nI(\mathbf{V}; \mathbf{Y})}$, and (b) follows from the fact that conditioning 
cannot increase entropy.

We bound the second term in \eqref{eq:secrecy_condition1_case2} as follows:
\begin{align*}
H(\mathbf{Z}^n|\mathbf{U}^n,\mathbf{V}^n) 
  &= H(\mathbf{Z}^n|\mathbf{U}^n)\\
  &= \sum_{\mathbf{u}^n} Pr(\mathbf{U}^n=\mathbf{u}^n)
H(\mathbf{Z}^n|\mathbf{U}^n=\mathbf{u}^n)\\
  &\stackrel{(a)}{=}  \sum_{\mathbf{u}^n} \PP(\mathbf{U}^n=\mathbf{u}^n)
         \sum_{\mu \in {\mathcal U}} N(\mu|\mathbf{u}^n)
H(\mathbf{Z}|\mathbf{U}=\mu)\\
  &\stackrel{(b)}{\geq} \sum_{\mathbf{u}^n} \PP(\mathbf{U}^n=\mathbf{u}^n)
         \sum_{\mu \in {\mathcal U}} n(\PP(\mathbf{U}=\mu) - \epsilon)
H(\mathbf{Z}|\mathbf{U}=\mu)\\
  &= \sum_{\mathbf{u}^n} \PP(\mathbf{U}^n=\mathbf{u}^n)
         (n H(\mathbf{Z}|\mathbf{U}) - n\epsilon_{14})\\
  &= nH(\mathbf{Z}|\mathbf{U}) - n\epsilon_{14},
\end{align*}
where (a) follows from the memoryless nature of the virtual channel from
$\mathbf{U}$ to $\mathbf{Z}$ and $N(\mu|\mathbf{u}^n)$ counts the number of times $\mu$
appears in the codeword $\mathbf{u}^n$, and (b) follows from the fact that
all the $\mathbf{u}^n$ codewords belong to $\typical_\epsilon^{\ast (n)}$.

The third term can be bounded by using Fano's inequality and the fact that
Eve can recover the $\mathbf{U}^n$ codeword with a probability of error
$\epsilon$ when she has access to $M$ and $\mathbf{V}^n$ in addition to her
observation $\mathbf{Z}^n$.
\begin{align*}
	H(\mathbf{U}^n | M, K, \mathbf{Z}^n, \mathbf{V}^n) \leq 1 + n \cdot
\epsilon \cdot I(\mathbf{U};\mathbf{Z} | \mathbf{V}) = n\epsilon_{15}.
\end{align*}

Finally, to bound the fourth term, let $T$ be an indicator random variable
which takes on the value 1 when $(\mathbf{V}^n,\mathbf{Z}^n) \in
\typical_\epsilon^{\ast (n)}$ and 0 otherwise.
\begin{align}
H(\mathbf{Z}^n|\mathbf{V}^n) 
 &\leq H(\mathbf{Z}^n, T |\mathbf{V}^n) \notag\\
 &\leq 1 + H(\mathbf{Z}^n|\mathbf{V}^n, T=1)\PP(T=1) +
   n\log|{\mathcal Z}|\PP(T=0). \label{eq:secrecycondition1term4_case2}
\end{align}
But
\[ \PP(T=0)=\PP((\mathbf{V}^n,\mathbf{Z}^n) \notin \typical_\epsilon^{\ast (n)}) \leq
\epsilon_{16}.\]
Furthermore, we have
\begin{align*}
H(\mathbf{Z}^n|\mathbf{V}^n, T=1) 
  &= \sum_{\mathbf{v}^n} \PP(\mathbf{V}^n=\mathbf{v}^n|T=1) 
                         H(\mathbf{Z}^n|\mathbf{V}^n=\mathbf{v}^n, T=1)\\
  &\stackrel{(a)}{\leq} \sum_{\mathbf{v}^n} \PP(\mathbf{V}^n=\mathbf{v}^n|T=1)
                      \log |\typical_\epsilon^{\ast(n)}(p_{\mathbf{Z}|\mathbf{V}}|\mathbf{v}^n)|\\
  &\leq \sum_{\mathbf{v}^n} \PP(\mathbf{V}^n=\mathbf{v}^n|T=1)
                      (nH(\mathbf{Z}|\mathbf{V}) + n\epsilon)\\
  &= nH(\mathbf{Z}|\mathbf{V}) + n\epsilon,
\end{align*}
where in (a) we used $|\typical_\epsilon^{\ast(n)}(p_{\mathbf{Z}|\mathbf{V}}|\mathbf{v}^n)|$ to
denote the size of the set of all $\mathbf{z}^n$ such that
$(\mathbf{z}^n,\mathbf{v}^n)\in \typical_\epsilon^{\ast(n)}$. Thus,
\eqref{eq:secrecycondition1term4_case2} becomes
\begin{align*}
H(\mathbf{Z}^n|\mathbf{V}^n) \leq nH(\mathbf{Z}|\mathbf{V}) + n \epsilon_{17}.
\end{align*}

Hence, we may conclude from \eqref{eq:secrecy_condition1_case2} that
\begin{align*}
\frac{1}{n}H(M, K|\mathbf{Z}^n) &\geq I(\mathbf{U};\mathbf{Y}|\mathbf{V}) +
H(\mathbf{Z}|\mathbf{U}) - H(\mathbf{Z}|\mathbf{V}) + \epsilon_{18}\\
  &= I(\mathbf{U};\mathbf{Y}|\mathbf{V}) -
I(\mathbf{U};\mathbf{Z}|\mathbf{V}) + \epsilon_{18}\\
  &= R_{\sf SM} + R_{\sf SK}+ \epsilon_{18}.
\end{align*}
Thus we have shown the secrecy condition.

\noindent{\em Proof of Uniformity Condition}.\\
Note that from \eqref{eq:prob_unif_bound}, we have that 
	\begin{align*}	
		H(K) &= H(\mathbf{V}^n, M, K, \Phi_{\sf sub-index}) - H(\mathbf{V}^n, M, \Phi_{\sf sub-index} | K) \\
		&\stackrel{(a)}{\geq}   H(\mathbf{V}^n, M, K, \Phi_{\sf sub-index}) - (I(\mathbf{V}; \mathbf{Y}) + n R_{\sf SM} + I(\mathbf{U};\mathbf{Z} | \mathbf{V})) \\
		&\stackrel{(b)}{\geq} H(\mathbf{V}^n, M, K, \Phi_{\sf sub-index} | S^n \in E) \cdot \PP (S^n \in E) \notag \\
		& \qquad - (I(\mathbf{V}; \mathbf{Y})+ n R_{\sf SM} + I(\mathbf{U};\mathbf{Z} | \mathbf{V})) \\ 
		&\stackrel{(c)}{\geq} n I(\mathbf{U};\mathbf{Y}) - (I(\mathbf{V}; \mathbf{Y}) + n R_{\sf SM} + I(\mathbf{U};\mathbf{Z} | \mathbf{V})) - n \epsilon_{13} \\
		&= R_{\sf SK} - n \epsilon_{13}~.
	\end{align*}
where (a) follows since $\mathbf{V}^n$ is drawn from a codebook with no more than $2^{n I(\mathbf{V}; \mathbf{Y})}$ elements, $M$ has less than $2^{n R_{\sf SM} }$ elements, and $ \Phi_{\sf sub-index}$ has no more than $2^{nI(\mathbf{U};\mathbf{Z} | \mathbf{V})}$ elements; (b) since conditional entropy is less than 
or equal to entropy; and (c) from the lower bound in \eqref{eq:prob_unif_bound}. Since $\epsilon_{13} \to 0$ as $\epsilon \to 0$, we satisfy the uniformity condition.

\section{Proof of Theorem \ref{thm:outerbounds}}\label{app:outerbounds}

The achievability follows directly from Theorem~\ref{thm:achievability} by
setting the auxiliary random variables as follows.
\begin{align*}
V_1 &= (X_F,X_R),\\
V_2 &= (V_{2,F},X_R),\\
U_1 &= U_{1,F}.
\end{align*}

It is easy to see that this satisfies the Markov conditions on the
auxiliary random variables. Substituting these in the expression in
Theorem~\ref{thm:achievability} shows the achievability. The interpretation
is that, we have ignored the reversely degraded source component, and the
reversely degraded channel is used purely as a channel for public
communication.

To show the converse, let $J$ and $J^\prime$ be independent random
variables both uniformly distributed over $\{1,2,\ldots,n\}$ and
independent of all other random variables. To get the first condition
(ignoring $o(n)$ terms)
\begin{align*}
n(I(X_{F,J};Y_{F,J} )+I(X_{R,J};Y_{R,J}) )  &\geq nI({\bf X}_{J};{\bf Y}_{J} )\\
 &\geq nI({\bf X}_{J};{\bf Y}_{J} | J)\\
 &\geq I({\bf X}^n;{\bf Y}^n)\\
 &\stackrel{\text{(a)}}{=} I({\bf X}^n;{\bf Y}^n,Z_{F}^n)\\
 &= I(M,K,{\bf S}_{A}^n,{\bf X}^n;{\bf Y}^n,Z_{F}^n)\\
 &\geq I(M,K,{\bf S}_{A}^n;{\bf Y}^n,Z_{F}^n)\\
 &\geq I(M,K,{\bf S}_{A}^n;{\bf Y}^n,Z_{F}^n) - I({\bf S}_{B}^n,{\bf S}_{E}^n;{\bf Y}^n,Z_{F}^n)\\
 &\stackrel{\text{(b)}}{=}
  I(M,K,{\bf S}_{A}^n;{\bf Y}^n,Z_{F}^n|{\bf S}_{B}^n,{\bf S}_{E}^n)\\
 &= I(M;{\bf Y}^n,Z_F^n|{\bf S}_{B}^n,{\bf S}_{E}^n) 
     + I(K,{\bf S}_{A}^n;{\bf Y}^n,Z_F^n|{\bf S}_{B}^n,{\bf S}_{E}^n,M)\\
 &\stackrel{\text{(c)}}{=} H(M|{\bf S}_{B}^n,{\bf S}_{E}^n) 
     + I(K,{\bf S}_{A}^n;{\bf Y}^n,Z_F^n|{\bf S}_{B}^n,{\bf S}_{E}^n,M)\\
 &= H(M) + I(K,{\bf S}_{A}^n;{\bf Y}^n,Z_F^n|{\bf S}_{B}^n,{\bf S}_{E}^n,M)\\
 &= nR_{\sf SM} +
    I(K,{\bf S}_{A}^n;{\bf Y}^n,Z_F^n|{\bf S}_{B}^n,{\bf S}_{E}^n,M)\\
\end{align*}
where (a) is due to the sub-channel $F$ to Eve being degraded w.r.t.
the channel to Bob, (b) is because $({\bf S}_{B}^n,{\bf S}_{E}^n) -
{\bf S}_{A}^n - (M,K,{\bf Y}^n,Z_F^n)$ is a Markov chain, and (c) follows
from Fano's inequality which gives $H(M|{\bf Y}^n,{\bf S}_{B}^n)=o(n)$.
Now, to bound the second term, we write
\begin{align*}
I(K,{\bf S}_{A}^n;&{\bf Y}^n,Z_F^n|{\bf S}_{B}^n,{\bf S}_{E}^n,M)\\
 &= H({\bf Y}^n,Z_F^n|{\bf S}_{B}^n,{\bf S}_{E}^n,M) -
    H({\bf Y}^n,Z_F^n|K,M,{\bf S}_{A}^n,{\bf S}_{B}^n,{\bf S}_{E}^n)\\
&\geq H({\bf Y}^n,Z_F^n|{\bf S}_{B}^n,{\bf S}_{E}^n,M) - 
      H(K,{\bf Y}^n,Z_F^n|{\bf S}_{A}^n,{\bf S}_{B}^n,{\bf S}_{E}^n,M)\\
&\stackrel{\text{(a)}}{=} H(K,{\bf Y}^n,Z_F^n|{\bf S}_{B}^n,{\bf S}_{E}^n,M) -
                H(K,{\bf Y}^n,Z_F^n|{\bf S}_{A}^n,{\bf S}_{B}^n,{\bf S}_{E}^n,M)\\
&= I(K,{\bf Y}^n,Z_F^n;{\bf S}_{A}^n|{\bf S}_{B}^n,{\bf S}_{E}^n,M)\\
&\stackrel{\text{(b)}}{=}
   I(M,K,{\bf Y}^n,Z_F^n;{\bf S}_{A}^n|{\bf S}_{B}^n,{\bf S}_{E}^n)\\
&\geq
   I(M,K,{\bf Y}^n,Z_F^n;S_{A,F}^n|S_{A,R}^n,{\bf S}_{B}^n,{\bf S}_{E}^n)\\
&= I(M,K,{\bf Y}^n,Z_F^n;S_{A,F}^n|S_{A,R}^n,S_{B,F}^n,S_{E,F}^n)\\
&= \sum_{i=1}^n I(M,K,{\bf Y}^n,Z_F^n;S_{A,F,i}|
                  S_{A,F}^{i-1},S_{A,R}^n,S_{B,F}^n,S_{E,F}^n)\\
&\geq \sum_{i=1}^n I(M,K,{\bf Y}^n,Z_F^n;S_{A,F,i}|
                     S_{A,R}^n,S_{B,F}^n,S_{E,F}^n)\\
&= \sum_{i=1}^n
  I(M,K,{\bf Y}^n,Z_F^n,S_{B,F,\tilde{i}},S_{E,F,\tilde{i}},S_{A,R}^n;
    S_{A,F,i}|S_{B,F,i},S_{E,F,i})\\
&= nI(M,K,{\bf Y}^n,Z_F^n,S_{B,F,\tilde{J}'},S_{E,F,\tilde{J}'},S_{A,R}^n;
      S_{A,F,J'}|S_{B,F,J'},S_{E,F,J'},J')\\
&\stackrel{\text{(c)}}{=}
 nI(M,K,{\bf Y}^n,Z_F^n,S_{B,F,\tilde{J}'},S_{E,F,\tilde{J}'},S_{A,R}^n,J';
      S_{A,F,J'}|S_{B,F,J'},S_{E,F,J'})\\
&= n I(U_{1,F};S_{A,F,J'}|S_{B,F,J'},S_{E,F,J'})\\
&\stackrel{\text{(d)}}{=} n I(U_{1,F};S_{A,F,J'}|S_{B,F,J'})\\
&\stackrel{\text{(e)}}{=} n(I(U_{1,F};S_{A,F,J'}) - I(U_{1,F};S_{B,F,J'})),
\end{align*}
where we define
$S_{B,F,\tilde{i}}\defineqq(S_{B,F}^{i-1},S_{B,F,i+1}^n)$,
$S_{E,F,\tilde{i}}\defineqq(S_{E,F}^{i-1},S_{E,F,i+1}^n)$, and\\
$U_{1,F}\defineqq(M,K,{\bf Y}^n,Z_F^n,S_{B,F,\tilde{J'}},S_{E,F,\tilde{J'}},
S_{A,R}^n,J')$. Note that 
(a) follows from Fano's inequality which implies that
$H(K|{\bf Y}^n,{\bf S}_{B}^n)=o(n)$, (b) follows 
the independence of $M$ from $({\bf S}_A^n,{\bf S}_B^n,{\bf S}_C^n)$. To see
(c), note that $({\bf S}_{A,J'},{\bf S}_{B,J'},{\bf S}_{E,J'})$ has the same
joint distribution as $({\bf S}_A,{\bf S}_B,{\bf S}_E)$. This equivalence of
joint distributions together with the fact that $U_{1,F}$ does indeed
satisfy the Markov condition $U_{1,F} - {\bf S}_{A,J'} - ({\bf
S}_{B,J'},{\bf S}_{E,J'})$ implies that $U_{1,F} - S_{A,F,J'} - S_{B,F,J'}
- S_{E,F,J'}$ is a Markov chain, which gives us (d) and (e). To get condition 2, 
\begin{align*}
n(R_{\sf SK}+R_{\sf SM}) 
&\leq I(M,K;{\bf Y}^n,Z_F^n,{\bf S}_{B}^n,{\bf S}_{E}^n)\\
&\stackrel{\text{(a)}}{=} I(M,K;{\bf Y}^n,Z_F^n,{\bf S}_{B}^n,{\bf S}_{E}^n) 
                           - I(M,K;{\bf Z}^n,{\bf S}_{E}^n)\\
&\stackrel{\text{(b)}}{=} I(M,K;{\bf Y}^n,Z_F^n,{\bf S}_{B}^n,{\bf S}_{E}^n) 
                           - I(M,K;{\bf Z}^n,Y_{R}^n,{\bf S}_{E}^n)\\
&\leq I(M,K;{\bf Y}^n,Z_F^n,{\bf S}_{B}^n,{\bf S}_{E}^n)
                           - I(M,K;Z_{F}^n,Y_{R}^n,{\bf S}_{E}^n)\\
&\stackrel{\text{(c)}}{=} I(M,K;Y_F^n,S_{B,F}^n|Y_{R}^n,Z_F^n,{\bf S}_{E}^n)\\
&= I(M,K;Y_F^n|Y_{R}^n,Z_F^n,{\bf S}_{E}^n) +
   I(M,K;S_{B,F}^n|{\bf Y}^n,Z_F^n,{\bf S}_{E}^n)\\
&\leq I(M,K,{\bf S}_{E}^n,Y_{R}^n,X_F^n;Y_F^n|Z_F^n) 
      + I(M,K;S_{B,F}^n|{\bf Y}^n,Z_F^n,{\bf S}_{E}^n)\\
&= I(X_F^n;Y_F^n|Z_F^n) + 
   \sum_{i=1}^n I(M,K;S_{B,F,i}|{\bf Y}^n,Z_F^n,S_{B,F}^{i-1},{\bf S}_{E}^n)\\
&= H(Y_F^n|Z_F^n) - \sum_{i=1}^n H(Y_{F,i}|X_{F,i},Z_{F,i}) +
   \sum_{i=1}^n I(M,K;S_{B,F,i}|{\bf Y}^n,Z_F^n,S_{B,F}^{i-1},{\bf S}_{E}^n)\\
&\leq \sum_{i=1}^n H(Y_{F,i}|Z_{F,i}) -
      \sum_{i=1}^n H(Y_{F,i}|X_{F,i},Z_{F,i}) + 
      \sum_{i=1}^n I(M,K,{\bf Y}^n,Z_F^n,S_{B,F,\tilde{i}},S_{E,F,\tilde{i}},
                   S_{A,R}^n;S_{B,F,i}|{\bf S}_{E,i})\\
&= nI(X_{F,J};Y_{F,J}|Z_{F,J},J) + 
      nI(M,K,{\bf Y}^n,Z_F^n,S_{B,F,\tilde{J}'},S_{E,F,\tilde{J}'},
                   S_{A,R}^n;S_{B,F,J'}|S_{E,F,J'},J')\\
&\leq nI(X_{F,J};Y_{F,J}|Z_{F,J},J) + nI(U_{1,F};S_{B,F,J'}|S_{E,F,J'})\\
&\stackrel{\text{(d)}}{=} n(I(X_{F,J};Y_{F,J}|V_{2,F})-I(X_{F,J};Z_{F,J}|V_{2,F})) +
     n(I(U_{1,F};S_{B,F,J'}) - I(U_{1,F};S_{E,F,J'})) 
\end{align*}
where $V_{2,F}\defineqq J$, (a) follows from the hypothesis
$I(M,K;{\bf Z}^n,{\bf S}_{E}^n)=o(n)$,
(b) from the fact that 
$I(M, K; Y_{R}^n | {\bf S}_{E}^n, {\bf Z}^n)$ $=
0$, which we show below, (c) from the Markov chain 
$(M,K,{\bf Y}^n,Z_F^n,{\bf S}_{A}^n) - S_{E,R}^n - S_{B,R}^n$, and (d) from the degradation
of the source component $F$ and the sub-channel $F$.
\begin{align*}
	0 = I({\bf S}_{A}^n,M, K; Y_{R}^n | {\bf Z}^n) 
	\stackrel{\text{(a)}}{=} 
              I({\bf S}_{E}^n, {\bf S}_{A}^n,M, K; Y_{R}^n | {\bf Z}^n) 
	&\geq I(M, K; Y_{R}^n | {\bf S}_{E}^n, {\bf Z}^n) ~,
\end{align*}
where (a) follows from the Markov chain ${\bf S}_{E}^n - ({\bf S}_{A}^n,M, K) 
- {\bf Z}^n - Y_{R}^n$. By non-negativity of mutual information, 
$I(M, K; Y_{R}^n | {\bf S}_{E}^n, {\bf Z}^n) = 0$ as claimed above.

Thus, we have shown that if $(R_1,R_2)\in\CapR$, then there must
exist random variables $U_{1,F}$ and $V_{2,F}$ jointly distributed with
$X_F,Y_F,Z_F,$ $X_R,Y_R,Z_R,$ $S_{A,F},S_{B,F},S_{E,F},$
$S_{A,R},S_{B,R},S_{E,R}$ such that their joint distribution is of the
following form
\begin{align*}  p_{S_{A,F},S_{B,F},S_{E,F}} p_{S_{A,R},S_{B,R},S_{E,R}}
p_{U_{1,F}|S_{A,F},S_{A,R}} p_{V_{2,F},X_F} p_{Y_F,Z_F|X_F} p_{X_R}
p_{Y_R,Z_R|X_R},
\end{align*}
and
\begin{align}
R_{\sf SM}&\leq I(X_F,Y_F ) + I(X_{R};Y_{R} ) - (I(U_{1,F};S_{A,F}) -
I(U_{1,F};S_{B,F})),\label{eq:appconverse1}\\
R_{\sf SK}+R_{\sf SM}&\leq I(X_F;Y_F|V_{2,F}) - I(X_F;Z_F|V_{2,F}) +
                            I(U_{1,F};S_{B,F}) - I(U_{1,F};S_{E,F}).
                    \label{eq:appconverse2}
\end{align}
The form of the right hand sides above further allows us to assert that the
$U_{1,F}$ above may be independent of $S_{A,R}$, i.e., it is enough to
consider joint distributions of the form
\begin{align}  p_{S_{A,F},S_{B,F},S_{E,F}} p_{S_{A,R},S_{B,R},S_{E,R}}
p_{U_{1,F}|S_{A,F}} p_{V_{2,F},X_F} p_{Y_F,Z_F|X_F} p_{X_R}
p_{Y_R,Z_R|X_R}. \label{eq:appconverse3}
\end{align}
This completes the proof.

\paragraph{Bandwidth mismatch} Suppose there is a bandwidth mismatch of
$m_S$ source symbols for every $m_C$ channel symbols. Then, for a
blocklength $n$ of channel symbols, we have $n_S\defineqq\lfloor{nm_S/m_C}\rfloor$ source
symbols. The only modification we need to make to the converse is to set
$J'$ to be uniformly distributed over $\{1,2,\ldots,n_S\}$. It is
straightforward to verify that the arguments carry over to the bandwidth
mismatch setting.

\paragraph{Stochastically degraded case} Theorem~\ref{thm:outerbounds}
also holds when the channels and sources are only stochastically degraded.
Achievability follows directly from Theorem~\ref{thm:achievability}. For
the converse, let us recall the definition of stochastic degradation. For
the source component $F$ made up of $S_{A,F},S_{B,F},S_{E,F}$, stochastic
degradation means that there is a conditional distribution
$p_{\tilde{S}_{E,F}|S_{A,F},S_{B,F}}$ such that we may define a random
variable $\tilde{S}_{E,F}$ jointly distributed with $S_{A,F},S_{B,F}$ which
satisfies (i) $S_{A,F}-S_{B,F}-\tilde{S}_{E,F}$ is a Markov chain, and (ii)
$(S_{A,F},\tilde{S}_{E,F})$ has the same joint distribution as
$(S_{A,F},S_{E,F})$.  Without loss of generality, we may assume that the
joint distribution of $S_{A,F},S_{B,F},S_{E,F},\tilde{S}_{E,F}$ follows
$p_{S_{A,F},S_{B,F}}$$p_{S_{E,F}|S_{A,F},S_{B,F}}$$p_{\tilde{S}_{E,F}|S_{A,F},S_{B,F}}$.
Similarly, for the subchannel $F$ given by $p_{Y_F,Z_F|X_F}$, there is a
conditional distribution $p_{Y_F,\tilde{Z}_F|X_F}$ such that $X_F - Y_F -
\tilde{Z}_F$ is a Markov chain and $p_{\tilde{Z}_F|X_F}$ is the same as
$p_{Z_F|X_F}$.  Again, without loss of generality, the conditional
distribution of $Y_F,Z_F,\tilde{Z}_F$ conditional on $X$ may be assumed to
follow $p_{Y_F|X_F}p_{Z_F|X_F,Y_F}p_{\tilde{Z}_F|X_F,Y_F}$.  Similarly, we
have $\tilde{S}_{B,R}$ and $\tilde{Y}_R$. Let $\tilde{\bf
S}_E=(\tilde{S}_{E,F},{S}_{E,R})$, $\tilde{\bf S}_B =
(S_{B,F},\tilde{S}_{B,R})$, $\tilde{\bf Y}=(Y_F,\tilde{Y}_R)$, and
$\tilde{\bf Z}=(\tilde{Z}_F,{Z}_R)$.

Notice, first of all, that for any coding
scheme $I(M,K;{\bf S}_E^n,{\bf Z}^n)$ only depends on the joint distribution
of random variables available at Alice and Eve. By the definition of
stochastic degradation, this implies that 
\[I(M,K;\tilde{\bf S}_E^n,\tilde{\bf Z}^n) = I(M,K;{\bf S}_E^n,{\bf Z}^n).\]
Hence, secrecy condition also applies for a {\em dummy} Eve who receives
$(\tilde{\bf S}_E^n,\tilde{\bf Z}^n)$ instead of the actual observations of Eve,
i.e., $I(M,K;\tilde{\bf S}^n,\tilde{\bf
Z}^n) = o(n).$ Similarly, the probability of decoding error for Bob is only a
function of the joint distribution of random variables at Alice and Bob
which is again preserved if we consider a {\em dummy} Bob who receives
$(\tilde{\bf S}_B^n,\tilde{\bf Z}^n)$. Hence, we may now repeat our
converse arguments for the setup with dummy Bob and dummy Eve who have
physically degraded sources and channels. We can verify that if $(R_1,R_2)\in\CapR$,
our converse proof in fact implies the existence of $U_{1,F}$ and $V_{2,F}$
which satisfy
\begin{align}
R_{\sf SM}&\leq I(X_F,Y_F ) + I(X_{R};\tilde{Y}_{R} ) - (I(U_{1,F};S_{A,F}) -
I(U_{1,F};S_{B,F})),\label{eq:stochconverse1}\\
R_{\sf SK}+R_{\sf SM}&\leq I(X_F;Y_F|V_{2,F}) - I(X_F;\tilde{Z}_F|V_{2,F}) +
                            I(U_{1,F};S_{B,F}) - I(U_{1,F};\tilde{S}_{E,F}).
                    \label{eq:stochconverse2}
\end{align}
with joint distributions of the form
\begin{align}  
p_{S_{A,F},S_{B,F}}p_{S_{E,F}|S_{A,F},S_{B,F}}p_{\tilde{S}_{E,F}|S_{A,F},S_{B,F}}
p_{S_{A,R},S_{B,R},\tilde{S}_{B,R},S_{E,R}}
p_{U_{1,F}|S_{A,F}} p_{V_{2,F},X_F} p_{Y_F,Z_F,\tilde{Z}_F|X_F} p_{X_R}
p_{Y_R,\tilde{Y}_R,Z_R|X_R}. \label{eq:stochconverse3}
\end{align}
Using the fact that $p_{\tilde{Y}_R|X}=p_{Y_R|X}$ we can replace
$\tilde{Y}_R$ in \eqref{eq:stochconverse1}. Similarly, using the fact that
$p_{\tilde{Z}_F|X_F}=p_{Z_F|X_F}$ (which by \eqref{eq:stochconverse3} 
implies that $p_{V_{2,F},X_F,\tilde{Z}_F}=p_{V_{2,F},X_F,Z_F}$) we may
replace $\tilde{Z}_F$ in \eqref{eq:stochconverse2} by $Z_F$. By a similar
argument, we may also replace $\tilde{S}_{E,F}$ in
\eqref{eq:stochconverse2} by $S_{E,F}$. Finally, by marginalizing away the dummy
variables in \eqref{eq:stochconverse3} we have the result for
stochastically degraded case as well.

\section{Proof of
Proposition~\ref{prop:Gaussianexample}}\label{app:Gaussianexample}

While we stated the
Theorems~\ref{thm:achievability}~and~\ref{thm:outerbounds} only for finite
alphabets, the results can be extended to continuous alphabets.  We note
that the scalar Gaussian problem satisfies the conditions of
Theorem~\ref{thm:outerbounds} (along with Remark~1 following it).

Observe that in the notation of Theorem~2, $S_{A,F}=S_A$ and $S_{B,F}=S_B$.
Further, $S_{A,R},S_{B,R},S_{E,F}$, and $S_{E,R}$ are absent (assumed to be
constants). When, $\text{SNR}_\text{Eve} \geq \text{SNR}_\text{Bob}$, we
have $X_R=X, Y_R=Y$, and $Z_R=Z$, and the forwardly degraded sub-channel is
absent (again, we may take the random variables of this sub-channel to be
constants). When $\text{SNR}_\text{Bob}\geq\text{SNR}_\text{Eve}$, we have
$X_F=X, Y_F=Y$, and $Z_F=Z$ and the reversely degraded sub-channel is
absent. Hence, from Theorem~2, $\CapR$ is given by the union of
$\tilde{\mathcal R}(p)$ over all joint distributions $p$.
Also, $\tilde{\mathcal R}(p)$ is described by
\begin{align}
R_{\sf SM} &\leq I(X_F;Y_F) + I(X_{R};Y_{R}) - I(U_1;S_{A}|S_{B}),
\label{eq:uppercase1}\\
R_{\sf SK}  + R_{\sf SM} &\leq I(X_F;Y_F|V_2) - I(X_F;Z_F|V_2) +
                           I(U_1;S_{B}).\label{eq:uppercase2}
\end{align}
When specialized to the Gaussian case above, it is easy to see that
\begin{align*}
I(X_F;Y_F)+I(X_R;Y_R)&\leq C_Y,\text{ and}\\
I(X_F;Y_F|V_2)-I(X_F;Z_F|V_2)&\leq [C_Y-C_Z]_+,
\end{align*}
where $C_Y=\frac{1}{2}\log(1+\text{SNR}_\text{Bob})$ and
$C_Z=\frac{1}{2}\log(1+\text{SNR}_\text{Eve})$. These bounds are
simultaneously achieved when $p$ is such that $V_2$ is a constant and $X$
is Gaussian of variance $\text{SNR}_\text{Bob}$. Hence, we may rewrite, the
conditions above as 
\begin{align}
R_{\sf SM} &\leq C_Y - I(U_1;S_{A}) + I(U_1;S_{B}),
\label{eq:Gaussuppercase1}\\
R_{\sf SK}  + R_{\sf SM} &\leq [C_Y-C_Z]_+ +
              I(U_1;S_{B}).\label{eq:Gaussuppercase2}
\end{align}

Now we show outerbounds to the above $\tilde{\mathcal R}(p)$ which match
the two conditions in proposition~\ref{prop:Gaussianexample}. It will also
become clear that a jointly Gaussian choice for $p$ in fact achieves
these outerbound thus completing the proof. We first derive an upperbound on
$R_{\sf SM}$ which matches the first condition in
proposition~\ref{prop:Gaussianexample}. From the two inequalities
\eqref{eq:Gaussuppercase1} and \eqref{eq:Gaussuppercase2} above, we have
\begin{align}
R_{\sf SM} &\leq C_Y - I(U_1;S_{A}) + I(U_1;S_{B}),
\label{eq:GaussSMuppercase1}\\
R_{\sf SM} &\leq [C_Y-C_Z]_+ +
              I(U_1;S_{B}).\label{eq:GaussSMuppercase2}
\end{align}
Using entropy power inequality,
\begin{align*}		 
\exp(2h(S_B|U)) &\geq \exp(2h(S_A|U)) + \exp(2h(N_\text{source}))\\
\end{align*}
Using this in \eqref{eq:GaussSMuppercase1}, we may write
\begin{align*}
\exp(2R_{\sf SM})
 &\leq \exp(2(C_Y+I(U_1;S_B)-h(S_A))) \left(\exp(2(h(S_B)-I(U_1;S_B))) -
              \exp(2h(N_\text{source})) \right)\\
 &= \exp(2(C_Y-h(S_A)+h(S_B))) - \exp(2(C_Y-h(S_A)+h(N_\text{source})))
                                 \exp(2I(U_1;S_B))\\
 &\stackrel{(a)}{\leq} \exp(2(C_Y-h(S_A)+h(S_B))) -
  \exp(2R_{\sf SM})\exp(2(C_Y-[C_Y-C_Z]_+-h(S_A)+h(N_\text{source}))),
\end{align*}
where (a) results from \eqref{eq:GaussSMuppercase2}.
Rearranging, we have
\begin{align*}
R_{\sf SM} &\leq \frac{\exp(2(C_Y-h(S_A)+h(S_B)))}
                      {1+\exp(2(C_Y-[C_Y-C_Z]_+-h(S_A)+h(N_\text{source})))}\\
           &= \frac{(1+\text{SNR}_\text{Bob})(1+\text{SNR}_\text{src})}
  {1+\text{SNR}_\text{src}+\min(\text{SNR}_\text{Bob},\text{SNR}_\text{Eve})}
\end{align*}
which is the first condition in proposition~\ref{prop:Gaussianexample}.
Now let us fix $R_{\sf SM}$ such that it satisfies this condition. 
Let us rewrite \eqref{eq:Gaussuppercase1} as follows
\begin{align*}
h(S_A|U) &\geq( R_{\sf SM} - C_Y + h(S_A) - h(S_B)) + h(S_B|U).
\end{align*}
Entropy power inequality implies that
\begin{align*}
 \exp(2h(S_B|U)) &\geq \exp(2h(S_A|U)) + \exp(2h(N_\text{source}))\\
&\geq \exp(2(R_{\sf SM}-C_Y+h(S_A)-h(S_B)))\exp(2h(S_B|U)) + 1.
\end{align*}
Since \[R_{\sf SM} \leq
           \frac{(1+\text{SNR}_\text{Bob})(1+\text{SNR}_\text{src})}
  {1+\text{SNR}_\text{src}+\min(\text{SNR}_\text{Bob},\text{SNR}_\text{Eve})}
\leq \frac{1}{2}\log\frac{(1+\text{SNR}_\text{Bob})(1+\text{SNR}_\text{src})}{\text{SNR}_\text{src}}
=C_Y-h(S_A)+h(S_B)
,\]
we have
\[ \exp(2h(S_B|U)) \geq \frac{1}{1-\exp(2(R_{\sf
SM}-C_Y+h(S_A)-h(S_B)))}.\]
From \eqref{eq:Gaussuppercase2},
\begin{align*}
\exp(2R_{\sf SK}) &\leq \exp(2([C_Y-C_Z]_+ + h(S_B) - h(S_B|U) - R_{\sf
SM}))\\
 &\leq \exp(2([C_Y-C_Z]_+ + h(S_B) - R_{\sf SM}))
           (1-\exp(2(R_{\sf SM}-C_Y+h(S_A)-h(S_B))))\\
 &\leq \exp(2([C_Y-C_Z]_+-C_Y))
           (\exp(2(C_Y+h(S_B)-R_{\sf SM})) - \exp(2h(S_A)))
\end{align*}
which evaluates to the second condition required. The inequalities used above
are tight under a Gaussian choice for the auxiliary random variable which
proves the achievability.

\end{appendices}

\bibliographystyle{IEEEtran}

\end{document}